\documentclass[11pt]{article}

\usepackage{amsfonts,amssymb,color,cancel,mathtools,setspace,tabularx}
 \usepackage{sidecap}
  \usepackage{amsmath}
\usepackage{graphicx}
\usepackage{amssymb}
\usepackage{pifont}
\usepackage{ulem}
\usepackage{framed}
\newcommand{\xmark}{\ding{55}}%

\usepackage{hyperref}
\hypersetup{
 colorlinks=true, 
    linktoc=page,     
    linkcolor=blue,
}
\newcommand{\bea}{\begin{eqnarray}}
\newcommand{\eea}{\end{eqnarray}}
\newcommand{\be}{\begin{equation}}
\newcommand{\ee}{\end{equation}}

\newcommand{\st}[1]{\ifmmode\text{\sout{\ensuremath{#1}}}\else\sout{#1}\fi} 

\makeatletter

\begin{document}


\title{\LARGE Emergent Black Hole Thermodynamics from Monodromy}

\author{
{\large Alex Chanson}$^{1}$ \footnote{a.chanson@aggiemail.usu.edu}, {\large Jacob Ciafre}$^{1}$ \footnote{jakeciafre@outlook.com} and {\large Maria J. Rodriguez}$^{1,2}$ \footnote{maria.rodriguez@usu.edu, maria.rodriguez@csic.es}, \\
\\
$^{1}${\small  Department of Physics, Utah State University,}\\ {\small 4415 Old Main Hill Road, UT 84322, USA}\\
\\
\\ 
 $^{2}${\small  Instituto de F\' isica Te\' orica UAM/CSIC,}\\ { \small Universidad Aut\'  onoma de Madrid, Cantoblanco, 28049 Madrid, Spain} \\
\\
\\
 }


\maketitle

\abstract

We argue that the equations of motion of quantum field theories in curved backgrounds encode new fundamental black hole thermodynamic relations. We define new entropy variation relations. These `emerge' through the monodromies that capture the infinitesimal changes in the black hole background produced by the field excitations. This raises the possibility of new thermodynamic relations defined as independent sums involving entropies, temperatures and angular velocities defined at every black hole horizon. We present explicit results for the sum of all horizon entropy variations for general rotating black holes, both in asymptotically flat and asymptotically anti-de Sitter spacetimes in four and higher dimensions. The expressions are universal, and in most cases add up to zero. We also find that these thermodynamic summation relations apply in theories involving multi-charge black holes.

\newpage

\tableofcontents
\newpage

\section{Introduction}

Quantum field theory in curved backgrounds is expected to provide an accurate description of quantum phenomena occurring in the early universe, as well as near -- and inside of -- black holes.
The simplest form of the equations of motion are embodied in the Klein-Gordon (KG) equation, which describes a massless scalar field in a curved background. Surprisingly, the KG-equations not only encode the analytic structure of the background geometries, but also more fundamental black hole thermodynamic relations. In this paper we will describe the emergence of a new thermodynamic identity from the KG-equation, the sum of the variational horizon entropies, that will allow us to identify new universal thermodynamic relations for black holes.

 The discovery of the thermodynamic behavior of black holes has given rise to most of our present physical insights into the quantum nature in the strong field regime.
Most famously, in 1973, Bardeen, Carter, and Hawking \cite{Bardeen:1973gs} provided a general proof of the first laws of thermodynamics of black holes 
%
with the intensive quantities defined at the black hole event horizon $r_+$. Yet, eternal black hole solutions contain a much richer geometrical horizon structure which includes Cauchy horizons. More recently, it was shown in \cite{Castro:2012av} that there is a universal ``geometrical first law of thermodynamics'' for a Cauchy horizons, the so called inner event-horizon $r_-$ of black holes. And, while physically unobservable, the fact that the inner horizon obeys standard thermodynamic relations is an indication of the validity of the CFT description \cite{Castro:2013lba,Castro:2013kea}. This additional horizon structure appears to play an important role in the precise description of generic microstates \cite{Cvetic:1997uw}. It is natural then to inquire whether analogously a geometrical thermodynamic laws holds for every (real or complex valued) Cauchy horizon in more general classes of black holes.

In this paper, we consider the geometrical relation of thermodynamics for every black hole horizon:
\begin{equation}\label{firstlaw}
dE=T_i \,dS_i+\sum_k \Omega^{(k)}_i \, dJ_{(k)}+\sum_l \Phi^l_i\, dQ_l\,,
\end{equation}
%
Here $S_i$ is the entropy, and the extensive quantities -- the total mass or energy $E$, the angular momenta $J_{(k)}$, and the total charge $Q_l$ -- are the Komar charges. The corresponding intensive quantities  -- the temperature $T_i$, the angular velocities $\Omega^{(k)}_i$, and electromagnetic potential $\Phi^l_i$--  are defined at each black hole
horizon.  The $i$-subscript represents the $r_i$ black hole event or Cauchy horizon where the thermodynamic quantity was defined, the index  $k=1,2,...,[\frac{d-1}{2}]$ represents the independent planes of rotation in a $d$-spacetime and $l$ the number of electric/magnetic charges.  An explicit examination of a fairly extensive number of black hole solutions reveals that the geometrical law of thermodynamics $(\ref{firstlaw})$ is indeed universal for all black hole horizons. 

Remarkably, we find that the geometrical first law of thermodynamics $(\ref{firstlaw})$ for every black hole horizon emerges from the KG-equation (via monodromies). We argue that the variations with respect to the conserved charges of the black hole entropies $\delta S_i$ are directly proportional to the the monodromies $\alpha_i$ around each horizon:
\begin{equation}\label{entropymon}
\delta S_i= \frac{(d-2)}{4}\,\mathcal{A}_{d-2}\,{\alpha}_i
\end{equation}
where 
$\mathcal{A}_{d-2}=2 \pi^{(d-1) / 2} / \Gamma\left((d-1) / 2\right)$
is the area of a unit $(d-2)$-sphere. The entropy variation relation (\ref{entropymon}) follows from a rather remarkable extension of the ideas in \cite{Castro:2012av} for the black hole event horizon, which is recovered when e.g. $d = 4$ for the Kerr black hole solution
\begin{equation}
\label{alphaentropy}\delta S_+=4 \pi \, \tilde{\alpha}_+\,,\qquad \text{where} \,\,\, \tilde{\alpha}_+=\frac{(\omega-\Omega_+ m)}{4\pi\,T_+}\,.
\end{equation}
The scalar field excitations produce infinitesimal changes in the black hole mass $\omega=\delta M$  and angular momentum $m=\delta J$; these associations reproduce \eqref{firstlaw} for the Kerr black hole. Having verified (\ref{firstlaw}) explicitly for all horizons, we can confirm that the monodromies have an important physical interpretation, namely they represent the variation in the entropy (\ref{entropymon}). We find that these infinitesimal relations do not imply global identities, such as Smarr relations.

Critically, the KG-equation can be reduced to a Fuchsian-type radial anstatz. This equation encodes the monodromy coefficient at infinity $\alpha_{\infty}$ of points at $r=\infty$. Employing a similar argument as in (\ref{entropymon}), we can therefore also define a new monodromy/entropy variation relation at infinity, which is $\delta S_{\infty}=\frac{(d-2)}{4}\, \mathcal{A}_{d-2} ((\mathcal{K}-1)-\alpha_{\infty})$ with a constant parameter $\mathcal{K}$.

Subsequently 
we find another remarkable thermodynamic feature for black hole solutions emerging from the KG-equation: we find that the summation of all monodromies -- generally, a Fuchs(-type) relation -- defines a new thermodynamic relation of the entropies, namely: 
\bea \label{sumentropy}
\sum_i \delta S_i =\delta S_{\infty}\,,
\eea
We have explicitly verified this relation for an extensive list of black holes solutions. These include  Schwarzschild, Kerr, $d$-dimensional Schwarzschild and Myers-Perry (MP) \cite{Myers:1986un}, BTZ \cite{Banados:1992wn}, $d$-dimensional Schwarzschild and Kerr-(A)dS \cite{Carter:1968ks, Hawking:1998kw,Gibbons:2004js}, Reissner-Nordstrom \cite{Caldarelli:1999xj}, Kerr-Newman \cite{Newman:1965tw}, Kerr-Newman-AdS, $d$-dimensional RN-(A)dS,  $d=5$ min. gauged SUGRA  \cite{Chong:2005hr} and $d=6$ gauged SUGRA \cite{Chow:2008ip} black hole solutions. 

In all cases, except for Kerr and Kerr-Newman  \footnote{The Kerr black hole solution does partially break the condition for the sum of the entropies being independent of the extensive quantities. Namely, $\sum_i S_i=4\pi M^2$, thus $\sum_i \frac{1}{T_i}\ne0$. Further details can be found in Section \ref{section:Kerr}.}, $S_{\infty}$ seems independent of the extensive quantities for all black holes. Therefore, $\delta S_{\infty}=0$ and the relation $(\ref{sumentropy})$ leads to 
\begin{equation}\label{conditions0}
\delta\left(\sum_i S_i\right)=0\,,
\end{equation}
and the following new universal relations for black holes:
\begin{equation}\label{conditions1}
\sum_i \frac{1}{T_i}=0\,,\qquad \sum_i \frac{\Omega^{(k)}_i}{T_i}=0\,\,.
\end{equation}
 It is also possible to verify, by considering charged black hole solutions that
\begin{equation}\label{conditions2}
\sum_i \frac{\Phi_i^l}{T_i}=0\,.
\end{equation}
Our results are summarized in Table \ref{table:1}.\\

\begin{table}[h!]
\centering
\begin{tabular}{||c |c|c| c| c||} 
 \hline
  {\small  Black Hole} & $\sum_i \delta S_i =\delta S_{\infty}$ &   {\small  $ \sum_i T^{-1}_i=0$ } &  {\small  $\sum_i \Omega^a_i/T_i=0$ } & {\small  $\sum_i \Phi^a_i/T_i=0$}   \\ [0.5ex] 
  \hline\hline
   Schwarzschild&\checkmark & \xmark & - & - \\ 
 \hline
Kerr \cite{Kerr:1963ud} &\checkmark &\xmark & \checkmark & -  \\ 
\hline
{\small Reissner-Nordstrom (RN)} & \checkmark &\xmark& - & \xmark \\
 \hline
 Kerr-Newman (KN) \cite{Newman:1965tw} &\checkmark& \xmark & \checkmark & \xmark \\
 \hline
 Schwarzschild $d>4$ \cite{Tangherlini:1963bw} &\checkmark& \checkmark & -& - \\ 
\hline
Myers-Perry \cite{Myers:1986un} &\checkmark& \checkmark & \checkmark & -  \\  
\hline
BTZ \cite{Banados:1992wn} &\checkmark& \checkmark & \checkmark & -  \\
  \hline
   Schw-(A)dS &\checkmark& \checkmark & - & - \\ 
 \hline
 Kerr-(A)dS  \cite{Carter:1968ks} &\checkmark& \checkmark & \checkmark & -  \\
 \hline
  Schw-(A)dS$_d$ &\checkmark& \checkmark & -& - \\ 
  \hline
Kerr-(A)dS$_d$ \cite{Hawking:1998kw,Gibbons:2004uw}&\checkmark& \checkmark & \checkmark & - \\ 
   \hline
 RN$_d$\cite{Caldarelli:1999xj}&\checkmark& \checkmark & - & \checkmark \\ 
 \hline
 RN-(A)dS &\checkmark& \checkmark & - & \checkmark \\
 \hline
RN-(A)dS$_d$&\checkmark& \checkmark & - & \checkmark \\ 
\hline
 KN-(A)dS \cite{Caldarelli:1999xj}&\checkmark& \checkmark & \checkmark & \checkmark \\
 \hline
 $5d$ gauged SUGRA \cite{Chong:2005hr} &\checkmark& \checkmark & \checkmark & \checkmark \\
 \hline
$6d$ gauged SUGRA  \cite{Chow:2008ip} &\checkmark& \checkmark & \checkmark & \checkmark \\
\hline
\end{tabular}
\caption{This table summarizes new thermodynamic relations for a copious number of black hole solutions in $d \ge 3$ space-time dimensions. The results include the confirmation of the emergent relation for the sum of the entropy variations of each black hole horizon $\sum_i  \delta S_i =\delta S_{\infty}$ as well as new thermodynamic relations. These new relationships, jointly with the properties of the $\sum_i S_i$, are a byproduct of the geometrical relation of thermodynamics for all black hole horizons (\ref{firstlaw}). The \checkmark \, represent the thermodynamic relations that are fulfilled, while the \xmark \, signals those that fail to obey the relations. The horizon radii $r_i$ are defined by the radial function $\Delta(r_i)=0$.}
\label{table:1}
\end{table}

The paper proceeds as follows.  Section \ref{sec:Overview} includes an overview of the KG-equation, our results for the monodromies, and the Fuchs(-type) relation for all black holes listed in Table \ref{table:1}. Section \ref{sec:Scoords} contains the explicit physical parameters for general asymptotically flat or AdS black holes solutions -- which are in some cases neutral, charged or rotating -- in $d$-spacetime dimensions. These are defined at every black hole horizon and individually shown to be consistent with the first law of thermodynamics for all black hole horizons. In each case we also asses the relation (\ref{conditions0})-(\ref{conditions2}) employing purely thermodynamical quantities. In section \ref{sec:proof}, we prove the new relations (\ref{conditions1}) and (\ref{conditions2}) that follow from the first law of thermodynamics and the properties of the $\sum_i S_i$. Finally, in Section \ref{sec:discussion}, concluding remarks are presented.

%


\section{Overview}
\label{sec:Overview}

One can explore the geometry of a neutral black hole considering small perturbations of the background. One of the simplest possibilities is a minimally coupled scalar, i.e. a
massless scalar field that satisfies the KG-equation:
\bea\label{KGeq}
\frac{1}{\sqrt{-g}} \partial_{\mu}(\sqrt{-g} \, g^{\mu\nu} \partial_{\nu} \Phi)=0\,,
\eea
The solutions to this equation can be presented in such a way that the Killing symmetries deriving from stationarity, and the axial symmetries are satisfied. Then the wave function can be written:
\bea\label{eq:sol}
\Phi(t,r,\theta,\phi_k)=\exp^{-i \,\omega \, t+\sum _k i \, m_k \phi_k} T(\theta) \, R(r)\,,
\eea
by means of the time coordinate $t$, the radial coordinate $r$, the polar coordinate $\theta$, together with $[(d-1)/2]$ azimuthal angular coordinates $\phi_k$. 
From the black hole backgrounds (given in  e.g.  section \ref{sec:monthermo}) it is straightforward to write out the equation (\ref{KGeq}) explicitly. For black hole solutions to Einstein's equation of General Relativity (GR) $R_{\mu\nu}=\Lambda \, g_{\mu\nu}$ with  $\Lambda =-(d-1)/L^2$, in Boyer-Lindquist coordinates, the wave equation can be reduced in the radial $r$-plane to a second order differential equation (ODE) with singularities $r_1,...,r_\mathcal{K}$ and the point at infinity $r=\infty$. It is useful to define a ``evenness'' integer $\epsilon= (d-1) \mod 2$ which is $1$ for even $d$ and $0$ for odd $d$
to express 
\bea
\mathcal{K}=d + 1 - \epsilon\,. 
\eea
For black holes \footnote{We focus on the non-extremal black holes in this paper where $r_i\ne r_j$. The extremal black hole cases will be studied elsewhere.} , while $r_1,...,r_\mathcal{K}$ are regular singular points, $r=\infty$ can either be a regular or an irregular singular point. For instance, in d-spacetime dimensions we find that the point $r=\infty$ is an irregular singular point in asymptotically flat black holes ($\Lambda=0$) and is regular for asymptotically AdS black holes ($\Lambda\ne0$) -- see Appendix \ref{app:IrregularSingPoints} for details. This distinction between regular and irregular singular points might seem artificial, but their implications for $R(r)$ are starkly different. Regular singular points have series expansion around say $r_i$ of the form
\bea\label{formalsol}
R(r)= (r-r_i)^{ i \alpha_{i}} [1+O(r-r_i)]\,.
\eea
whereas, the asymptotic expansion for the solutions  e.g., around $r =\infty$, one must also include exponential factors in the series expansion
\bea\label{formalsol2}
R(r)=e^{ i\omega r} r^{ i \lambda_{\infty}-(d-2)/2} [1+O(r^{-1})]\,.
\eea
where $ \lambda_{\infty}$ is the eigenvalue of the formal monodromy, that we will refer to as {\it fake monodromy}.\\

Having identified the singular points, we turn to the Fuchs relation, which describes a direct relationship between the asymptotic exponents $\alpha_i$( i.e. the {\it monodromies}) of formal series solutions (\ref{formalsol}) of Fuchsian linear differential equations. We argue that the radial ODE for AdS black holes has all regular singularities (including $r=\infty$) and is therefore a Fuchsian type equation with a corresponding Fuchs relation.  As we will now show, the Fuchs relation relating the sum of the monodromies around each regular singular point of the ODE is exactly the relation of the sum of the variation of the black hole entropies defined at each horizon (\ref{sumentropy}). Furthermore, we will establish as a limiting case a Fuchs-type relation for the asymptotically flat black holes. The new Fuchs-type relation also gives account to the sum of the variation of the black hole entropies defined at each horizon that is found from purely thermodynamical relations. In this sense, we find a new {\it emergent black hole thermodynamic relation from monodromies relations}.



Understanding the Fuchs relation will suffice to illustrate the essence of our approach to the sum of entropy variations, so we focus first on asymptotically AdS black holes in section \ref{subsec:fuchs}, discussing the added complications of irregular singular points for asymptotically flat black holes in section \ref{subsec:fuchstype}.

\subsection{ODE with all regular singular points: Fuchsian equation}
\label{subsec:fuchs}

Here we give the fundamentals of our approach  to the sum of entropy variations for asymptotically AdS black holes in all dimensions. We begin by defining the second-order Fuchsian equation with all $\mathcal{K}$ regular singularities at $r=r_i$, where $i=1,2,...,\mathcal{K}$, and at $\infty$. This is given by
\bea\label{eq:FuchsGen}
\frac{d^2 R}{dr^2}+p(r) \,\frac{d R}{dr}+q (r) \, R=0\,,
\eea
with rational functions of the form
\bea\label{eq:FuchsGenPQ}
p(r)=\sum_{i=1}^{\mathcal{K}}\frac{1-\alpha_i}{(r-r_i)}\,, \qquad q(r)=\sum_{i-1}^{\mathcal{K}}\frac{q_i}{(r-r_i)}.
\eea
A brief review can be found in \cite{DLMF}. This representation of equation (\ref{eq:FuchsGen}), that we will refer to as {\it frame} choice, corresponds to a setup for a Fuchsian equation with $\lim_{r\rightarrow r_i }q(r) (r-r_i)^2 =0$ where one of the two monodormy exponent parameters (around the regular singular point) is null. 
This particular frame gives us the most simple form for the monodormies to establishing the link with the entropy variations as we explain here below. 
While other frame choices are possible, we found that alternative frames as defined in \cite{Poole} give the same results with slightly more involved relations.

When the differential equation (\ref{eq:FuchsGen}) has regular singularities over $P_1(\mathbb{C})$, its exponents obey the so called Fuchs relation (\cite{Poole}, Chapter. V, p. 78):
\bea \label{eq:fuchs}
  \sum_{i=1}^{\mathcal{K}} \alpha_i =(\mathcal{K}-1) - \alpha_{\infty} \,.
\eea
Note that at $r=r_i$, the indicial equation is 
\bea\label{indicial}
\alpha_i \,  (\alpha_i-1)+ p_0 \, \alpha_i +q_0=0\,,
\eea
where
\bea
p_0=\lim_{r\rightarrow r_i} (r-r_i) \,p(r)\,,\qquad q_0=\lim_{r \rightarrow r_i} (r-r_i)^2  \, q(r)\,.
\eea
For $(\ref{eq:FuchsGen})$, one finds that $p_0=1-\alpha_i$ and $q_0=0$, so that the $\mathcal{K}$-non-trivial monodromy exponent parameters at the finite singularities $r_i$ are $\alpha_i$. The monodromy exponent parameter $\alpha_{\infty}$ relative to $r=\infty$ regular singular point is given by (\ref{formalsol}) transformed by $x=r^{-1}$ at $x=0$.

Having established our conventions, we proceed to present our general approach for AdS black holes in $d$-spacetime dimensions.
An examination of the radial equation derived from the KG equation in AdS black hole backgrounds in all dimensions shows that the ODE contains all regular singular points, including asymptotic infinity ($p_0,q_0$ are finite at every singular point). It is therefore a Fuchsian equation, which its parameters are subject to the Fuchs relation (\ref{eq:fuchs}). To implement the Fuchs relation we set the frame that simplifies the analysis by bringing the ODE to the form (\ref{eq:FuchsGen}). The parameter $\mathcal{K}$ corresponds to the total number of (event or Cauchy) horizons in the black hole solution.

%

Extending the ideas in \cite{Castro:2012av}, we argue that at every (regular singular point) black hole horizon in (\ref{eq:fuchs}) we can identify the monodromies with the entropy variations as in equation (\ref{entropymon}).
The Fuchs relation then encodes an emergent relation on the sum of entropy variations
\begin{equation}
 \sum_{i=1}^{\mathcal{K}} \alpha_i =(\mathcal{K}-1) - \alpha_{\infty}  \qquad \rightarrow  \qquad \sum_{i=1}^{\mathcal{K}} \delta S_i =\delta S_{\infty}\,,
\end{equation}
where we defined $\delta S_{\infty}=\frac{(d-2)}{4}\, \mathcal{A}_{d-2}  ((\mathcal{K}-1)-\alpha_{\infty})$. As we will illustrate explicitly in section \ref{sec:monthermo}, the variation $\delta S_{\infty}=0$ for asymptotically AdS black holes. Therefore, replacing these emergent relations between the entropy variations and monodromies, the Fuchs relation (\ref{eq:fuchs}) can be interpreted as the sum of the variation of entropies
\bea\label{eq:AdSBH}
 \boxed{ \sum_{i=1}^{\mathcal{K}}\delta S_i =0 } \, \qquad \text{for AdS-black holes}.
\eea
We verified this new emerging thermodynamic relation for the sum of the variations of entropies in section \ref{sec:Scoords} employing purely thermodynamic definitions. Having identified this relation, we were able to find and verify the sum of the entropies being independent of all intrinsic or extrinsic parameters
\bea
  \delta \left(\sum_{i=1}^{\mathcal{K}} S_i \right)=0
  \eea
which follows (\ref{eq:AdSBH}) and other universal relations  (\ref{conditions1}) for black holes such as:
\begin{equation}
\sum_{i=1}^{\mathcal{K}}\frac{1}{T_i}=0\,,\qquad \sum_{i=1}^{\mathcal{K}} \frac{\Omega^{(k)}_i}{T_i}=0\,\,.
\end{equation}
We present a proof for the latter relations in section \ref{sec:proof}.

\subsection{ODE with one irregular singular point: Fuchs-type relation}
\label{subsec:fuchstype}

In contrast with the black hole solutions in curved spacetimes, the radial equation derived from the KG equation in asymptotically flat black hole backgrounds everywhere regular except at asymptotic infinity, where it is irregular. This ODE is not given by a Fuchsian equation, yet we can still find a Fuchs-type relation that holds for asymptotically flat black holes. Asymptotically flat black hole solutions of GR result from taking a limit of vanishing cosmological constant $\Lambda \rightarrow 0$. Equivalently, since $\Lambda =-(d-1)/L^2$, the vanishing cosmological constant limit corresponds to $L \rightarrow \infty$. The Fuchs-type relation that we find for asymptotically flat spacetimes (with one irregular singularity at $r=\infty$) follows from taking the vanishing cosmological constant limit ($L\rightarrow \infty$ in our case) in equation (\ref{eq:fuchs}):
\bea 
\lim_{L\rightarrow\infty} \left(\sum_{i=1}^{\mathcal{K}} \alpha_i - (\mathcal{K}-1) + \alpha_{\infty} \right) = 0 \,,\\
 \boxed{ \sum_{i=1}^{\mathcal{K}-2} \alpha^{flat}_i -\lambda_{\infty}=0}\,.  \label{eq:fuchstype}
\eea
We will refer to (\ref{eq:fuchstype}) as the Fuchs-type relation for asymptotically flat black holes containing one irregular singular point at $\infty$. This relation results from $\mathcal{K}-2$ monodromies remaining finite
\bea \label{eq:Fuchtype}
\lim_{L\rightarrow \infty} \alpha_i =\alpha^{flat}_i  \,,
\eea
 around the $\mathcal{K}-2$ singularities that, in the $L\rightarrow \infty$ limit, remain (finite) regular singular points. The other two regular singular points converge to $r=\infty$ and, interestingly, the limiting sum of their corresponding monodromies is precisely the fake monodromy $\lambda_{\infty}$ as defined in (\ref{formalsol2}):
\bea 
 \lim_{L\rightarrow \infty} \alpha_{{\mathcal{K}}-1}+ \alpha_{\mathcal{K}}=-\lambda_{\infty}\,, 
\eea 
We can implement the relation (\ref{eq:fuchstype}) to asymptotically flat black holes (such as Kerr or Myers-Perry black hole solutions). The total number of (event or Cauchy) horizons in the black hole solution  in this case is $\bar{\mathcal{K}}=d -1 - \epsilon$ with $\epsilon =0,1$ respectively for odd or even d-dimensional spacetimes (or alternatively, in terms of the $\mathcal{K}$, $\bar{\mathcal{K}}=\mathcal{K}-2$).  As in the previous section, employing equation (\ref{entropymon}) relating the monodromies to the entropy variations at each black hole horizon\sout{s} (which are always regular singularities) we find through (\ref{entropymon}) that

\bea
\sum_{i=1}^{\mathcal{K}-2} \alpha^{flat}_i -\lambda_{\infty}=0 \qquad \rightarrow \qquad \sum_{i=1}^{\mathcal{K}-2} \delta S^{flat}_i = \delta S^{flat}_{\infty}\,,
\eea
where $\delta S^{flat}_{\infty}=\frac{(d-2)}{4}\, \mathcal{A}_{d-2} \lambda_{\infty}$. We therefore argue that a new universal thermodynamic relation involving the sum of the variation of entropies emerges from the Fuchs-type relation (\ref{eq:fuchstype}) for asymptotically flat black hole solutions with $\bar{\mathcal{K}}$ (event and Cauchy) horizons:
\bea \label{eq:sumflat}
 \boxed{ \sum_{i=1}^{\bar{\mathcal{K}}} \delta S^{flat}_i = \delta S^{flat}_{\infty}}\, \qquad \text{ for asymptotically flat black holes}.
\eea
As we will explicitly show in Appendix \ref{app:IrregularSingPoints}, the fake monodromy for asymptotically flat black holes is
\bea\label{eq:MoninftyAsymFlat}
&&\text{For Asympt. Flat}_4: \lambda_\infty\rightarrow 4 M \omega\,, \\
&&\text{For Asympt. Flat}_{d>4}: \lambda_\infty\rightarrow 0\,.
\eea
We verified in these cases that the new emerging thermodynamic relation for the sum of the variations of entropies (\ref{eq:sumflat}) employing solely the physical thermodynamic parameters  -- see section \ref{sec:Scoords} for details. A new set of thermodynamic relations  (\ref{conditions1}) also arises for asymptotically flat black holes with $\lambda_\infty = 0$ 
\begin{equation}
\sum_{i=1}^{\bar{\mathcal{K}}}\frac{1}{T_i}=0\,,\qquad \sum_{i=1}^{\bar{\mathcal{K}}} \frac{\Omega^{(k)}_i}{T_i}=0\,\,.
\end{equation}
that follow from (\ref{eq:sumflat}). Section \ref{sec:proof} contains further details and proofs.

\section{Monodromies and Emergent Thermodynamics}
\label{sec:monthermo}

This section focuses on the study of a scalar field in the d-dimensional (asymptotically AdS and flat) Schwarzschild and Kerr black hole background, but the methods are readily extendable to a broad class of physically relevant situations. We will first revisit the wave equation of the probes with particular emphasis on the machinery to compute the (finite and fake) monodromies.

\subsection{$AdS_d$ Schwarzschild Black Hole}
To setup our notation and conventions, we start by reviewing aspects of the geometry of d-dimensional Schwarzschild black hole with mass M. In Boyer-Lindquist coordinates, we have
\bea\label{eq:SchAdS}
ds^2=-\Delta \, dt^2+\frac{dr^2}{\Delta}+ r^2 \, d\Omega_{d-2}^2 \,,
\eea
where $\Delta=1-\frac{2M}{r^{d-3}}+\frac{r^2}{L^2}=(r^{d-3}L^2)^{-1}\prod_{i=1}^{\mathcal{K}-1}(r-r_i)$ with $\mathcal{K}=d + 1 - \epsilon$.

\subsubsection*{Wave Equation}

The Klein-Gordon equation for a massless scalar is (\ref{KGeq}) and using (\ref{eq:sol}), makes the equation separable. When the separation constant are set to zero, the radial equation for $R(r)$ is given by
\bea\label{eq:KGAdSS}
\frac{1}{r^{d-2}} \, \partial_r \left( r^{d-2} \Delta \, \partial_r R(r) \right) +\frac{\omega^2 R(r)}{\Delta}=0\,.
\eea
%
When choosing
\bea\label{frame}
R(r) =\prod_{i=1}^{\mathcal{K}-1} (r-r_i)^{-i\tilde{\alpha_i}} \tilde{R}(r)\,,\qquad
%
\text{with} \qquad
%
\tilde{\alpha_i}
=\frac{\omega}{\Delta'(r_i)}\,,
\eea
%
the ODE for $\tilde{R}(r)$ is a Fuchsian equation of the form (\ref{eq:FuchsGen}) with  (\ref{eq:FuchsGenPQ}) given by
\bea \label{eq:pq}
p(r)=\frac{2 \, r^{d-2} \Delta \, \partial_r P+P \, \partial_r(r^{d-2} \Delta)}{r^{d-2} \Delta \,P}\,,\qquad q(r)=\frac{\partial_r^2 P}{P}+\frac{\partial_r(r^{d-2} \Delta)  (\partial_r P)}{r^{d-2} \Delta \,P}+\frac{\omega^2}{\Delta^2},
\eea
where $P=\prod_{i=1}^{\mathcal{K}-1} (r-r_i)^{-i\tilde{\alpha_i}}$. In turn, the monodromy exponents $\{ \alpha_0, \alpha_i, \alpha_{\infty}\}$ with $i=1,2,...,\mathcal{K}-1$ in the formal series solutions (\ref{formalsol}) for $\tilde{R}(r)$ obey the Fuchs relation
\bea \label{eq:fuchs1}
 \sum_{i=0}^{\mathcal{K}} \alpha_i =(\mathcal{K}-1) - \alpha_{\infty}\,.
\eea

\subsubsection*{Monodromies and Regular Singular Points}
One can identify the nature of the singular points, for example, by following the steps in section \ref{sec:Overview} and appendix \ref{app:IrregularSingPoints}. As it turns out in this case, all singularities in the ODE for $\tilde{R}(r)$ are regular singular points.
 In addition to $\infty$, the $\mathcal{K}$ finite regular singularities are located at $r=\{0,r_i \}$. The monodromies, that we find via the indicial equation, are respectively
\bea\label{AdSmonodromies}
 \alpha_{\infty}=\mathcal{K}-1, \qquad \alpha_0=0, \qquad  \alpha_i=2\,\tilde{\alpha_i} \qquad \text{for} \qquad i=1,2,...,\mathcal{K}-1 \,,
  \eea

 An alternative way of determining the monodromy of the singularity at, say, $r=0$, would be to substitute a series expansion (\ref{formalsol}) and study the behavior of the ODE near these points.
And, similarly, substituting the series expansion (\ref{formalsol}) and changing $r\rightarrow x^{-1}$ in (\ref{eq:KGAdSS}) for $\tilde{R}(r)$ we find that the monodromy around $x=0$ ($r=\infty$).
%
 %
\subsubsection*{Sum of Entropy Variations}

Having computed the monodormies we can verify that these obey the Fuchs relation (\ref{eq:fuchs}). We find, replacing the relation (\ref{entropymon}) between the monodormies and the entropy variations 
in the Fuchs relation, that the sum of the monodromy parameters is equivalent to the sum of the variations of the entropies:
\bea
 \sum_{i=1}^{\mathcal{K}}{\alpha}_i =0\,,  \qquad \rightarrow \qquad  \sum_{i=1}^{\mathcal{K}}\delta S_i =0\,, \qquad \rightarrow \qquad   \delta \left(\sum_{i=1}^{\mathcal{K}} S_i \right)=0\,,
 \eea
where $S_i= (\mathcal{A}_{d-2}/4) \, r_i^{d-2}$ is the entropy computed at each horizon of the AdS$_d$ Schwarzschild black hole. We further verified this new entropy bound employing purely thermodynamic relations in section \ref{sec:Scoords}. In agreement with the previous result in $d=4$, equation (\ref{entropymon}) becomes $S_+=4\pi \,\tilde{\alpha_+}$ as found in \cite{Castro:2012av}.

\subsection{Schwarzschild Black Hole in $d$-dimensions}

We begin by briefly describing the geometry and radial part of the wave equation for an asymptotically flat $d$-dimensional Schwarzschild black hole describing a generic asymptotically flat static black hole with mass M. The linelement is of the form (\ref{eq:SchAdS}) where the function $\Delta\rightarrow \bar{\Delta}=1-\frac{2M}{r^{d-3}}=(r^{d-3})^{-1}\prod_{i=1}^{\bar{\mathcal{K}}-1}(r-r_i)$ and $\bar{\mathcal{K}}=d -1 - \epsilon$. 

\subsubsection*{Wave Equation}

The massless scalar Klein-Gordon equation in the background of a d-dimensional Schwarzschild black hole solution is separable. Employing the ansatz (\ref{eq:sol}), and setting the separation constant to zero, the radial equation for the function $R(r)$ becomes
\bea\label{eq:canonicalKGS}
\frac{1}{r^{d-2}} \, \partial_r \left( r^{d-2} \bar{\Delta} \, \partial_r R(r) \right) +\frac{\omega^2 R(r)}{\bar{\Delta}}=0\,.
\eea
%
%
As we did in the previous section, we choose a frame (\ref{frame}) with $\mathcal{K}\rightarrow \bar{\mathcal{K}}$ to find the corresponding functions $p(r)$ and $q(r)$ which become (\ref{eq:pq}) with $\Delta\rightarrow \bar{\Delta}$.

\subsubsection*{Fake Monodromies and Irregular Singular Points}

As it turns out in this case, the singularities in the ODE (\ref{eq:canonicalKGS}) are
 \bea
 r=\{ 0,(2M)^{1/(d-3)} e^{ \mathrm{i}(2 \pi\, (i-1))/(d-3)},\infty \}\qquad \text{where} \,\,\,  i=1,2,...,d-3\,.
 \eea 
 Employing the procedure described in Appendix \ref{app:IrregularSingPoints}, we find  that all but $\infty$ are regular singularities. A way of determining the monodromy of the regular points of the ODE 
would be to solve the indicial equation (\ref{indicial}). Following the previously described steps for regular singular points, we find that 
\bea
\alpha^{flat}_0=0 \,,\qquad \alpha^{flat}_i= \frac{ 2 \omega}{\bar{\Delta}'(r_i)}\,.
 \eea
 %
In order to compute the monodromy $\lambda_{\infty}$ around the irregular singular point $r=\infty$, we analyze the ODE
changing $r\rightarrow x^{-1}$. We present the details to compute the fake monodromy exponent $\lambda_{\infty}$ in Appendix \ref{app:IrregularSingPoints}. Our results are summarized in (\ref{eq:MoninftyAsymFlat}).
It is easily verified that these flat space-time monodormies $\{\alpha^{flat}_i, \lambda_{\infty}\}$ can also be found from those in AdS-Schwarzschild backgrounds in taking the flat space-time limit, $L\rightarrow \infty$ while keeping the other physical quantities fixed. For example in $d=4$, the non-vanishing monodoromies (\ref{AdSmonodromies}) yield 
\bea
\lim_{L\rightarrow\infty} \alpha_+=\alpha^{flat}_+\,,\qquad \lim_{L\rightarrow\infty}\alpha_2+\alpha_3=- 4 M \omega=- \lambda_{\infty}\,,
 \eea 
 
\subsubsection*{Sum of Entropy Variations}

In general, the monodromy exponents in ODEs that have an irregular singular point do not satisfy a Fuhcs relation. However in the present flat case, taking the sum of the monodromy coefficients  found in the preceding subsection leads to a relation which is consistent  with the Fuhcs-type relation. Plugging these expressions into (\ref{eq:fuchstype}) and employing the relation (\ref{entropymon})  between the monodormies and the entropy variations 
\bea
 \sum_{i=0}^{d-3} \alpha_i =\lambda_{\infty}\,,  \,\,\, \rightarrow \,\,\,  \sum_{i=0}^{d-3}\delta S_i =\delta S^{flat}_{\infty} \rightarrow \,\,\,   \delta \left(\sum_{i=0}^{d-3} S_i \right)=
 \begin{cases}
 8 \pi M \, \delta M  &d=4\\
0 &d >4
\end{cases}
 \eea
where $S_i= (\mathcal{A}_{d-2}/4) \, r_i^{d-2}$ is the entropy computed at each horizon of the $d$-dimensional Schwarzschild black hole. This new entropy bound is consistent with purely thermodynamic relations in section \ref{sec:Scoords}. 

\subsection{Kerr-AdS Black Hole}
\label{subsec:AdS-Kerr}

We start by reviewing aspects of a four dimensional AdS-Kerr black hole with mass M employing the notation in \cite{Castro:2013kea}. In $d=4$ AdS-spacetime, using Boyer-Lindquist coordinates, the corresponding line element is:
\bea
ds^2= \frac{\Sigma}{\Delta}dr^2 -\frac{\Delta}{\Sigma}(dt-\frac{a}{\Xi} \sin^2\theta d\phi)^2 +\frac{\Sigma}{\Delta_\theta} d\theta^2+\frac{\Delta_\theta}{\Sigma}\sin^2\theta\left(\frac{(r^2+a^2)}{\Xi}d\phi -a dt\right)^2
\eea
where
\bea
 \Delta_\theta &&= 1 - \frac{a^2}{l^2}\cos^2 \theta , \qquad{} \Xi= 1- \frac{a^2}{l^2} ,\qquad{} \Sigma=r^2+a^2\cos^2\theta, \\
&&\qquad{} \Delta=(r^2+a^2)(1+\frac{r^2}{L^2})-2Mr=\frac{1}{L^2}\prod\limits_{i=1}^{4}(r-r_i)
\eea
Further, we define the event horizons as the zeros of the $\Delta$ function: here, $\Delta$ has four roots two are real ($r_\pm$), two are imaginary ($r_{3,4}$). Therefore $\mathcal{K}_{AdSKerr}=4$. The analytic value of these roots, additional to particular properties of these roots, can be found in Appendix B.

\subsubsection*{Wave Equation}
Using the process illustrated in Appendix A and the results found in \cite{Castro:2013kea}, the differential equation for the radial ansatz is found to be:

\bea
\label{adsrad}\Big{[}\partial_r \Delta \partial_r +\sum_i \frac{(r_i^2+a^2)^2}{\Delta'(r_i)}\frac{(\omega-\Omega_i\, m)^2}{r-r_i} -L^2\Xi \,\omega^2 + \frac{a^2m^2}{L^2}\Big{]}R_{AdS}(r)=K_{l,AdS} \,R_{AdS}(r)
\eea
where
$
\Omega_i = \frac{a}{r_i^2+a^2}\Big{(} 1 + \frac{r_i}{L^2}\Big{)}\,.
$
Here $K_{l,AdS}$ is the angular coupling constant, that to simplify the calculations it will be set to zero here. For the form of the angular anstatz, see \cite{Castro:2013kea}.
Next, we follow the procedure of \eqref{frame}
\bea
R(r)=\prod\limits_{i=1}^4 (r-r_i)^{-i\tilde{\alpha}_i}\tilde{R}(r), \qquad{} \text{with} \qquad \tilde{\alpha}_i = \frac{r_i^2+a^2}{\Delta'(r_i)}(\omega-\Omega_i \, m)\,.
\eea
It can easily be verified that in this case 
\bea\label{eq:alphaRel}
\sum^4_{i=1} \tilde{\alpha}_i =0\,.
\eea
Then, with $\Delta^* = -L^2\Xi \, \omega^2 + \frac{a^2m^2}{L^2}$ after some algebra, \eqref{adsrad} becomes:
\bea
\Big{[}\partial_r^2 &+& \left(\sum\limits_{i=1}^{4}\frac{1- 2 i\,  \tilde{\alpha}_i}{r-r_i}\right)\partial_r  -\sum\limits_{i}^4 \sum\limits_{j\neq i}^4\frac{ \tilde{\alpha}_i(i+\tilde{\alpha}_j)}{(r-r_i)(r-r_j)} \nonumber \\
&+&\sum\limits_{i=1}^4\frac{\tilde{\alpha}_i^2}{(r-r_i)^2}\left(\frac{\Delta'(r_i)}{\prod\limits_{j\neq i}(r-r_j)} -1\right)+\frac{\Delta^*}{\Delta}\Big{]}\tilde{R}(r) =0 \label{AdSkerrgauged}
\eea
In the above expression is gauged so that, in the notation of \eqref{eq:FuchsGen}, $\lim\limits_{r\to r_i} q(r)(r-r_i)^2=0$. Further, here,  $\lim\limits_{r\to r_i}p(r)(r-r_i)=1- 2i \tilde{\alpha}_i$. 

\subsubsection*{Monodromies and Sum of Entropy Variations}
The indicial equation determining the monodromies $\alpha_i$ in the $\tilde{R}$-frame reads:
\bea\label{eq:monKerrAdS}
\alpha_i (\alpha_i-2\tilde{\alpha}_i)=0 \qquad \Rightarrow \alpha_i = \{ 0, 2\tilde{\alpha}_i \}=\left\{0,\frac{2(r_i^2+a^2)}{\Delta'(r_i)}(\omega-\Omega_i \, m)\right\}\,.
\eea
where to be consistent with the definition in (\ref{formalsol}) the coefficients $\alpha_i \in \mathbb{R}$.
By mapping $r\to x^{-1}$ and analyzing $x=0$, it can be proven independently that $\alpha_\infty = \{ - \sum\limits_{i=1}^{4}\tilde{\alpha}_i, 3- \sum\limits_{i=1}^{4}\tilde{\alpha}_i \}$. Considering the relation  (\ref{eq:alphaRel}) that was previously identified, we conclude that 
\bea
\alpha_\infty =\{0,3\}\,.
\eea
Our results are in agreement with $\alpha_{\infty}=\mathcal{K}_{AdSKerr}-1$ and the limiting AdS-Schwarschild results (\ref{AdSmonodromies}). It is worth noting that similar results hold in more exotic AdS (dS) spacetimes: for an analysis of dS-spacetime, see \cite{daCunha:2015fna}, for an analysis of NUT-spacetimes, see \cite{Novaes:2014lha}.  Using the same associations as \eqref{alphaentropy}:
\bea
 \sum_{i=1}^{4}{\alpha}_i =0\,,  \qquad \rightarrow \qquad  \sum_{i=1}^{4}\delta S_i =0\,, \qquad \rightarrow \qquad   \delta \left(\sum_{i=1}^{4} S_i \right)=0\,,
 \eea
where $S_i$ is the entropy computed at the $r_i$ horizon. Additionally, this new bound is re-verified using purely thermodynamic relations in section \ref{sec:Scoords}.

\subsection{Kerr Black Hole}

In this section we will consider a Kerr black hole with mass $M$ and angular momentum $J=Ma$. Using Boyer-Lindquist coordinates, for this asymptotically flat spacetime we find that:
\bea
ds^2= \frac{\Sigma}{\bar{\Delta}}dr^2 -\frac{\bar{\Delta}}{\Sigma}(dt-a \sin^2\theta d\phi)^2 +\Sigma \,d\theta^2 + \frac{\sin^2\theta}{\Sigma}\big{(}(r^2+a^2)d\phi -a dt\big{)}^2
\eea
where $\bar{\Delta}=r^2+a^2-2Mr=(r-r_-)(r-r_+)$ and $\Sigma=r^2+a^2\cos^2\theta$. As above, the event horizon radii are the zeros of the $\bar{\Delta}$ function; here, $r_{\pm} = M\pm\sqrt{M^2-a^2}$. Therefore, $\bar{\mathcal{K}}=2$. 

\subsubsection*{Wave Equation}
Using the process illustrated in Appendix A and \cite{Castro:2013lba}, the radial anstatz is found to be:
\bea
\Big{[}\partial_r \bar{\Delta} \partial_r +(r_+-r_-)\Big{(}\frac{\tilde{\alpha}^2_+}{r-r_+}-\frac{\tilde{\alpha}^2_-}{r-r_-}\Big{)} +\Delta^{*}\Big{]}R(r)=K_l R(r)\,,
\eea
where,
\bea
\tilde{\alpha}_\pm=\frac{\Delta'(r_{\pm})}{(r_{\pm}^2+a^2)}(\omega-\Omega_{\pm} m)\,,
\eea
the function $\Delta^* = (r^2+2M(r+2M))\, \omega^2$, 
and $K_l$ is the angular coupling constant. For the analysis of the angular anstatz, see \cite{Castro:2013lba}. We can again follow the transform of \eqref{frame} and $K_l=0$, leading to an analogue to \eqref{AdSkerrgauged}:
\bea
\Big{[}\partial_r^2 &+& \Big{(}\sum\limits_{i=1}^{2}\frac{1- 2 i\tilde{\alpha}_i}{r-r_i}\Big{)}\partial_r  -\frac{2\tilde{\alpha}_+\tilde{\alpha}_-+i(\tilde{\alpha}_++\tilde{\alpha}_-)}{\bar{\Delta}} \nonumber \\
&+&\sum\limits_{i=1, j \neq i}^2\frac{\tilde{\alpha}_i^2}{(r-r_i)^2}\Big{(}\frac{r_i-r_{j}}{r-r_j} -1\Big{)}+\frac{\bar{\Delta}^*}{\bar{\Delta}}\Big{]}\bar{R}(r) =0\,. \label{Kerrgauged}
\eea
As in \eqref{eq:FuchsGen}, $\lim\limits_{r\to r_{\pm}} q(r)(r-r_{\pm})^2=0$. Further, here,  $\lim\limits_{r\to r_{\pm}}p(r)(r-r_{\pm})=1- 2i \tilde{\alpha}_{\pm}$. Then, the indicial equation reads:
\bea 
{\alpha^{flat}_i}({\alpha_i}^{flat}-2\tilde{\alpha}_\pm)=0 \qquad \Rightarrow {\alpha^{flat}_i} = \{ 0, 2\tilde{\alpha}_i\}=\left\{0,\frac{2\Delta'(r_{\pm})}{(r_{\pm}^2+a^2)}(\omega-\Omega_{\pm} m)\right\}\label{eq:KerrMon}\,.
\eea
where $\alpha^{flat}_i \in \mathbb{R}$. Using the mapping $r\to x^{-1}$ and a Frobeinius expansion about $x=0$ it can be directly shown that $\lambda_\infty = 4M\omega$. Additionally, an analysis similar to that done in Appendix \ref{app:IrregularSingPoints} is also possible. For more analysis of the irregular singularity, including a discussion of it's unique scattering properties and the Stoke's phenomenon, see \cite{daCunha:2015ana}

\subsubsection*{Fake Monodromies and Sum of Entropy Variations}

From \eqref{eq:KerrMon}, we find the Fuchs-type relation $\alpha^{flat}_+ + \alpha^{flat}_- = 4M\omega$. It is also possible to show the monodromic sum evolves under $L\to \infty$ such as \eqref{eq:fuchstype}. In fact, the Kerr-AdS monodormies (\ref{eq:monKerrAdS}) in the limit yield
\bea
\lim_{L\rightarrow\infty} \alpha_\pm \to \alpha_\pm^{flat}, \qquad{} \lim_{L\rightarrow\infty}  {\alpha}_{3,4} \to -2M\omega \pm iL\omega
\eea
and the Fuchs relation
\bea
\lim_{L\rightarrow\infty} \left( \sum_{i=1}^{4}{\alpha}_i \right)=0 \qquad \rightarrow  \qquad \alpha^{flat}_+ + \alpha^{flat}_- -4M\omega=0\,.
\eea
Letting $\lambda_\infty := - \lim_{L\rightarrow\infty} (\alpha_3 + \alpha_4)=4M\omega$, and the identifications in (\ref{entropymon}) we find
\bea
  \alpha^{flat}_+ + \alpha^{flat}_-  =\lambda_\infty\, \,\, \rightarrow \qquad   \delta \left(S^{flat}_+ + S^{flat}_-\right)=\delta S^{flat}_\infty\,,
 \eea
where $S^{flat}_\pm$ is the entropy computed at the $r_\pm$ horizon. Additionally, this new bound is re-verified using purely thermodynamic relations in section \ref{sec:Scoords}. 

\section{Thermodynamic Identities of Black Holes}
\label{sec:Scoords}

We now turn to the study of the geometrical relation of thermodynamics for every black hole horizon (\ref{firstlaw}) and the novel thermodynamic identities of black holes (\ref{conditions0})-(\ref{conditions2}) from a purely thermodynamic perspective. Our focus is on thermodynamics properties of black hole solutions which have smooth horizons with spherical topologies. To establish a well rounded catalogue for the thermodynamic identities we shall present explicitly several examples in this section. Some of these results were previously assessed in literature. We indicate the references accordingly in each case.

In general these equations involve the physical parameters defined at each black hole horizon. It is convenient to define the intensive quantities employing employing the ADM formalism in which the line element is
\bea
ds^2 =-N^2 dt^2 + \gamma_{ab}(dx^a + N^a\,dt)(dx^b + N^b\,dt) ,
\eea
with $x^a$ spatial directions, and $N(x^a)$ and $N^b(x^a)$ the lapse function and the shift vector respectively. The intensive variables are intrinsic to each horizon that will be indicated by the subscripts.

Note that with this foliation of spacetime, the black hole horizons $r_i$ (real or complex) are at $N^2=0$. As in \cite{Astefanesei:2009wi}, the angular potentials and temperatures for each horizon are defined
\bea
\Omega^{(k)}_{i}=-\left. N^k \right|_{r_i},\qquad T_i=\left. \frac{1}{4\pi}\frac{(N^2)'}{\sqrt{g_{rr}N^2}}\right |_{r_i}
\eea
where $k=1,2,...,[\frac{d-1}{2}]$ represents the independent planes of rotation in a $d$-dimensions. With this definition for the temperature, when evaluated at the black hole event horizon $r_+$ we recover the Hawking temperature $T_+$.\\
In the ensuing analysis we will find the entropy sums over all horizons. In almost all cases the relation is independent of the extensive parameters which in turn implies the addition of the variations of the entropies to vanish. In all cases we find a perfect agreement with the corresponding results of the entropy and monodromy relations analyzed in previous sections. From each of these examples, new thermodynamic identities involving the sum of the intensive quantities are obtained.

\subsection{Kerr Black Hole}
\label{section:Kerr}

The radii function of a Kerr black hole \cite{Kerr:1963ud} of mass $M$ and angular momentum $J=M a$ satisfying $R_{\mu\nu}=0$ 
is defined by the function $\Delta(r)=r^2+a^2-2Mr$. The outer $r_+=M + \sqrt{M^{2}-a^{2}}$ and inner event horizons $r_-=M - \sqrt{M^{2}-a^{2}}$ are located at $\Delta(r_{\pm})=0$. The corresponding physical parameters are given by
\bea \label{KerrEntropy}
T_{\pm}&=&
\frac{\Delta'(r_{i})}{8\pi M r_{\pm}}\,,\qquad S_{\pm}={\pi (r_{\pm}^2+a^2)}\,,\qquad
\Omega_{\pm}=\frac{a}{r_\pm^2+a^2}\,.
\eea
corresponding to Hawking's temperature, the entropy, the angular momentum, and the angular velocity defined at 
black hole horizons. In this case we checked that the above physical parameters obey the relation (\ref{firstlaw}). The sum of the horizon areas is
\bea 
\sum_{i=1}^2 S_i =4\pi M^2\,,
\eea
The variation of this expression is
\bea
 \delta \left(\sum_{i=1}^2 S_i \right)=8 \pi M \, \delta M\,.
\eea
consistent with Section \ref{sec:monthermo}. Note also that
\bea  
\sum_{i=1}^2 \frac{1}{T_i}= 8\pi M\,,\qquad \sum_{i=1}^2 \frac{\Omega_i}{T_i}=0\,.
\eea
These can be viewed as a consequence of the first law of thermodynamics at each horizon $r_i$. A completely analogous story holds to all other black hole solutions as we proof in Section \ref{sec:proof}. A separate treatment of these cases are given in the following subsections.

\subsection{Kerr-AdS Black Hole}

The metric of the four dimensional Kerr-AdS black hole \cite{Carter:1968ks}, satisfying $R_{\mu\nu}=-3L^{-2}\,g_{\mu\nu}$ is asymtotic to $AdS_4$ in a rotating frame, with angular velocity $\Omega_{\infty}=-a L^{-2}$. The radii function is determined by $ \Delta=(r^2+a^2)(1+r^2L^{-2})-2Mr$. The outer event horizon and Cauchy horizons are located at $\Delta(r_{i})=0$ with $i=1,2,...,4$. The physical parameters corresponding to Hawking's temperature, the entropy, and the angular velocity of the horizon  (as measured in the asymptotically rotating frame) are given by
\bea
T_{i}&=&\frac{\Delta'(r_{i})}{4\pi(r_{i}^2+a^2)}\,,\qquad S_{i}=\frac{\pi (r_{i}^2+a^2)}{\Xi}\,,\qquad \Omega_{i}=\frac{a (1+r_+^2/L^2)}{r_+^2+a^2}\,.
\eea
where $\Xi=1-a^2/L^2$ and $\Omega_{i}$ is the angular velocity measured relative to a {\it rotating} observer at infinity. In \cite{Gibbons:2004ai} the physical mass $E$ and angular momentum $J$ of the AdS-Kerr black hole solution were computed at the boundary (infinity) via the Komar integrals
\bea
E=\frac{M}{\Xi^2}\,,\qquad J= E a\,.
\eea
It is straightforward to verify that these quantities obey the geometrical law of thermodynamics (\ref{firstlaw}). Further properties among the horizons of the Kerr-AdS black hole solution can be found in Appendix \ref{app:PropertiesAdSKerr}. The following relations then follow. The sum of the entropies at every horizon is independent of the physical parameters
\bea\label{eq:AdSSumS}
\sum^4_{i=1} S_i =-2\pi L^2\,,
\eea
and, also
\bea
\sum_{i=1}^4\frac{\Omega_i}{T_i}=0\,, \qquad \sum_{i=1}^4\frac{1}{T_i}=0\,,
\eea
In the study of this systems obeying the KG-equation, we found in Section \ref{subsec:AdS-Kerr} the thermodynamic relation (\ref{conditions0}) from the Fuchs relation. Given the sum of entropies relation (\ref{eq:AdSSumS}), one con compute its variation (keeping $L$ fixed) and thence obtain the associated thermodynamic relation $\delta \sum S_i =0$ consistent with previous results.
A completely analogous story holds for all the AdS-black holes in all dimensions. See sections \ref{subsec:BTZ} and \ref{subSec:AdSMP} for details.


\subsection{BTZ Black Hole}
\label{subsec:BTZ}\
The mass, angular momentum and entropy of the BTZ black hole  \cite{Banados:1992wn} are 
\bea
M=\frac{r_+^2+r_-^2}{L^2}\,,\qquad J=\frac{2 r_+ r_-}{L}\,,\qquad S_i=4\pi r_i\,,
\eea
and the Hawking temperature $T_i$ and angular velocity $\Omega_i$ are
 \bea
 T_i = \frac{r_+^2-r_-^2}{2 \pi  L^2 r_i} \,, \qquad\Omega_i = \frac{J}{2 r_i^2}\,,
 \eea
For this black hole solution, the roots of the radii equation
\bea
\frac{J^2}{4 r^2}+\frac{r^2}{L^2}-M=0
\eea
are found to be $r_i=\{\pm r_{+}, \pm r_-\}$ with $r_{\pm} =\sqrt{\frac{L (M L\pm \sqrt{M L^2-J^2 })}{2}}$.
Just as in the previous cases, we compute the sum of all entropies defined at each horizon $r_i$
\bea
\sum_{i=1}^4 S_i=0\,,
\eea
 Considering  the intensive quantities we further find 
\bea
\sum_{i=1}^4\frac{\Omega_i}{T_i}=0\,, \qquad \sum_{i=1}^4\frac{1}{T_i}=0\,.
\eea

\subsection{Higher Dimensional  Black Holes} 
In this paper we are interested also in higher dimensional black holes. The physical parameters for each $d\ge5$ black hole solution are described in the following subsections. Employing these quantities, we are able to establish new thermodynamic relations.

\subsubsection{Myers-Perry Black Hole in $d=5$}

The metric of the five dimensional rotating black hole, satisfying $R_{\mu\nu}=0$ was found in \cite{Myers:1986un}. The function
\bea
\Delta(r)=\frac{1}{r^2}(r^2+a_1^2)(r^2+a_2^2)-2M\,
\eea
defines the horizon radii located at $\Delta(r_{i})=0$ with $i=1,...,4$. While the four roots are real, only two are positive roots and correspond to the outer and inner event horizons. Hawking's temperature, the entropy, both angular momenta, the angular velocities of the horizon and physical mass parameters are respectively given by \footnote{In particular for $r_{\pm}$ we can write the temperature as $T_{\pm}=\frac{r_\pm^4-a_1^2 a_2^2}{2\pi r_\pm(r_\pm^2+a_1^2)(r_\pm^2+a_2^2)}$}
\bea\label{parametersMP}
T_{i}&=&\frac{r_i^2\Delta'(r_i)}{4\pi (r_i^2+a_1^2)(r_i^2+a_2^2)}\,,\qquad S_{i}=\frac{\pi^2 (r_{i}^2+a_1^2)(r_{i}^2+a_2^2)}{2 r_{i}}\,,\\
  J_{(k)}&=&\frac{\pi M a_k }{2}\,,\qquad \Omega^{(k)}_{i}=\frac{a_k}{r_i^2+a_k^2}\, \qquad E=\frac{3\pi}{4} M \nonumber
\eea
where \(k=1,2\).
%
The geometrical thermodynamical relation (\ref{firstlaw}) for the above quantities is satisfied as well as the following relations
\bea
\sum^4_{i=1} S_i =0\,,
\eea
and
\bea
\sum_{i=1}^4\frac{\Omega^{(k)}_i}{T_i}=0\,,\qquad  \sum_{i=1}^4\frac{1}{T_i}=0\,,
\eea
%
\subsubsection{Myers Perry Black Holes}
\label{subSec:MP}

The extension of Einstein's gravity for asymptotically flat rotating black holes in $d\ge 5$ space-time dimensions is shown in \cite{Hawking:1998kw, Gibbons:2004js}.
The $\mathcal{K}=d-1-\epsilon$ horizons can be found by the following radial equation.
\begin{equation}
\Delta(r)= r^{\epsilon-2} \prod_{k=1}^{[(d-1)/2]} \left(r^{2}+a_{k}^{2}\right)-2 M=0
\end{equation}
where $\epsilon= (d-1) \mod 2$.
As in \cite{Gibbons:2004ai} the entropy is defined by,
\begin{equation}
S_i=\frac{\mathcal{A}_{d-2}}{4 r_i^{1-\epsilon}}  \prod_{k=1}^{[(d-1)/2]} \left(r_{i}^{2}+a_{k}^{2}\right)
\end{equation}
The Hawking temperatures \(T_i\), angular velocities \(\Omega^{(k)}_i\), mass \(E\) and angular momenta \(J_{(k)}\) is given respectively by,
\begin{equation}
T_i
=\frac{1}{2 \pi} \left( r_{i} \sum_{k=1}^{[(d-1)/2]} \frac{1}{r_{i}^{2}+a_{k}^{2}}-\frac{2-\epsilon}{2 r_{i}}\right)\,,\qquad
\Omega^{(k)}_i=\frac{ a_{k}}{r_{i}^{2}+a_{k}^{2}}
\end{equation}
\begin{equation}
E=\frac{M \mathcal{A}_{d-2}}{4 \pi}\left(\frac{d-2}{2}\right)\,,\qquad
J_{(k)}=\frac{M a_{k} \mathcal{A}_{D-2}}{4 \pi }
\end{equation}
where $i=1,...,\mathcal{K}$ includes the outer and all Cauchy horizons.  For the above quantities, we find that geometrical thermodynamical relation (\ref{firstlaw}) is satisfied as well as the following relations
\bea
\sum_{i=1}^\mathcal{K} S_i =0
\eea
and
\bea
\sum_{i=1}^\mathcal{K} \frac{\Omega^{(k)}_i}{T_i}=0\,,\qquad  \sum_{i=1}^\mathcal{K} \frac{1}{T_i}=0\,
\eea
Some of the results in this subsection for the entropy sum, were derived in \cite{Xu:2013zpa}.

\subsubsection{Kerr-AdS Black Holes in $d\ge 5$}
\label{subSec:AdSMP}

The extension of Einstein's gravity for rotating black holes that are asymptotically AdS in $d\ge 5$ dimensions is shown in \cite{Hawking:1998kw, Gibbons:2004js}.
The outer and Cauchy horizons can be found by the following radial equation.
\begin{equation}
\Delta(r)= r^{\epsilon-2}\left(1+r^{2} L^{-2}\right) \prod_{k=1}^{[(d-1)/2]} \left(r^{2}+a_{k}^{2}\right)-2 M=0
\end{equation}
where $\epsilon= (d-1) \mod 2$.
As in \cite{Gibbons:2004ai} the entropy is defined by,
\begin{equation}
S_i=\frac{\mathcal{A}_{d-2}}{4 r_i^{1-\epsilon}}  \prod_{k=1}^{[(d-1)/2]} \frac{r_{i}^{2}+a_{k}^{2}}{1-a_k^2 L^{-2}}
\end{equation}
The Hawking temperatures \(T_i\), angular velocities \(\Omega^{(k)}_i\), mass \(E\) and angular momenta \(J_{(k)}\) are given respectively by,
\begin{equation}
T_i
=\frac{1}{2 \pi} \left( r_{i}\left(1+r_{i}^{2} L^{-2}\right) \sum_{k=1}^{[(d-1)/2]} \frac{1}{r_{i}^{2}+a_{k}^{2}}-\frac{2-\epsilon \,\left(1- r_{i}^{2} L^{-2}\right)}{2 r_{i}}\right)
\end{equation}
\begin{equation}
\Omega^{(k)}_i=\frac{\left(1+r_{i}^{2} L^{-2}\right) a_{k}}{r_{i}^{2}+a_{k}^{2}}
\end{equation}
\begin{equation}
E=\frac{M \mathcal{A}_{d-2}}{4 \pi\prod_{k=1}^{[(d-1)/2]}\left(1-a_k^2 L^{-2}\right) }\left(\sum_{i=1}^{[(d-1)/2]} \frac{1}{1-a_k^2 L^{-2}}-\frac{1-\epsilon}{2}\right)
\end{equation}
\begin{equation}J_{(k)}=\frac{M a_{k} \mathcal{A}_{d-2}}{4 \pi \left(1-a_k^2 L^{-2}\right) \prod_{k=1}^{[(d-1)/2]} \left(1-a_k^2 L^{-2}\right) }\end{equation}
where \(i\) includes the outer and all Cauchy horizons. The geometrical thermodynamical relation (\ref{firstlaw}) for the above quantities is satisfied as well as the following relations
\bea
\sum_{i=1}^\mathcal{K} S_i =\begin{cases}
0 &\text{d is odd}\\
\frac{L^{d-2}}{2} \, \mathcal{A}_{d-2} &\text{d is even}
\end{cases}
\eea
and
\bea
\sum_{i=1}^\mathcal{K} \frac{\Omega^{(k)}_i}{T_i}=0\,,\qquad  \sum_{i=1}^\mathcal{K} \frac{1}{T_i}=0\,
\eea
Some of the results in this subsection for the entropy sum, were derived in \cite{Xu:2013zpa,Liu:2016skp}.

\subsection{Charged Black Holes} 
We now turn to the study of the sum of entropies of charged charged holes. It is convenient to first define the relevant thermodynamic quantities. Given a black hole solution one can also compute new relations involving the intensive physical parameter such as (\ref{conditions1}) and (\ref{conditions2}). 

\subsubsection{Reissner-Nordstrom Black Hole} 
The metric of a four dimensional static, charged black hole in the usual Weyl coordinates is presented \cite{{Reissner:1916,Weyl:1917}} where the vanishing of the function
\begin{equation}
\Delta(r)=1-\frac{2  M}{r}+\frac{Q^2}{r^{2}}
\end{equation}
determines the horizon radii $\Delta(r_i)=0$. These are trivially,
\begin{equation}
r_{\pm}=M \pm \sqrt{M^{2}-Q^{2}}
\end{equation}
The entropy, Hawking's temperature and electric potential parameters are respectively given by
\begin{eqnarray}
S_{\pm}=\pi r_{\pm}^2 \,,\qquad
T_{\pm}=\frac{r_\pm ^2 - Q^2}{4 \pi r_\pm^3}  \,,\qquad
\Phi_{\pm}&=& \frac{Q}{r_{\pm}} \,.
\end{eqnarray}
The geometrical law of thermodynamics (\ref{firstlaw}) for the above quantities is satisfied as well as the following relations
\begin{equation}
\sum^2_{i=1} S_i =2 \pi \left( 2 M^2-Q^2 \right)
\end{equation}
and
\begin{equation}
\sum_{i=1}^2\frac{\Phi_i}{T_i}=4 \pi Q\,, \qquad  \sum_{i=1}^2\frac{1}{T_i}=8  \pi M
\end{equation}
\subsubsection{Reissner-Nordstrom-AdS  Black Hole}

The solution represnting a four dimensional static, charged black hole in AdS is presented in  \cite{{Reissner:1916,Weyl:1917}}. The function
\bea
\Delta(r)=1-\frac{2 M}{r}+\frac{Q^2}{r^2}+\frac{r^2}{3 L^2}
\eea
defines the horizon radii located at $\Delta(r_{i})=0$ with $i=1,...,4$. The entropy, Hawking's temperature and electric potential parameters are respectively given by
\bea
S_{i}=\pi r_i^2 \,,\qquad
T_{i}=  \frac{ 3 r_i^{4}L^{-2}+ r_i^2 -Q^{2}}{4 \pi   r_i^{3}}  \,,\qquad \Phi_{i}=  \frac{Q }{r_i} \,.
\eea
The geometrical law of thermodynamics (\ref{firstlaw}) for the above quantities is satisfied as well as the following relations
\bea
\sum^4_{i=1} S_i =-2 \pi L^2\,,
\eea
and
\bea
\sum_{i=1}^4\frac{\Phi_i}{T_i}=0\,, \qquad  \sum_{i=1}^4\frac{1}{T_i}=0\,,
\eea

\subsubsection{Reissner-Nordstrom Black Hole in $d\ge 5$}

A charged, non-rotating, asymptotically flat black hole in $d\ge 5$-dimensions is described in detail by \cite{Wei:2009zzc}. 
The radii function in this case can be written as
\begin{equation}
\Delta(r)=1-\frac{2 \mu}{r^{d-3}}+\frac{q^{2}}{r^{2(d-3)}}
\end{equation}
Integral to defining thermodynamic properties are the outer and inner most radii, 
\begin{equation}
r_{ \pm}^{d-3}=\mu \pm \sqrt{\mu^{2}-q^{2}}
\end{equation}
where,
\begin{eqnarray}\label{eq:defsRNHighDim}
\mu=\frac{8 \pi M}{\mathcal{A}_{d-2}(d-2)}  \,,\qquad
q^{2}=\frac{2 Q^{2}}{(d-2)(d-3)} \,.
\end{eqnarray}
Thus the ADM mass and charge can be constructed,
\begin{equation}
\begin{array}{l}{M=\frac{\mathcal{A}_{d-2}(d-2)}{16 \pi}\left(r_{+}^{d-3}+r_{-}^{d-3}\right)}  \,,\qquad {Q=\sqrt{\frac{(d-3)(d-2)}{2}}\left(r_{+} r_{-}\right)^{\frac{d-3}{2}}}\end{array}
\end{equation}
The entropy, Hawking temperature, and electric potential defined at the outer and inner most horizons can be generalized,
\begin{equation}
S_{i}=\frac{\mathcal{A}_{d-2}}{4} r_{i}^{d-2} \,,\qquad
T_{\pm}=\frac{d-3}{4 \pi r_{\pm}}\left[1-\left(\frac{r_{\mp}}{r_{\pm}}\right)^{d-3}\right] \,,\qquad
\Phi_{i}=\frac{\mathcal{A}_{d-2} Q}{4 \pi(d-3) r_{i}^{d-3}}\,.
\end{equation}
The geometrical relation of thermodynamics (\ref{firstlaw}) for the above quantities is satisfied as well as the following relations for \(d>4\)
\begin{equation}
\sum_{i=1}^{2 d-6} S_i=0
\end{equation}
and
\begin{equation}
\sum_{i=1}^{2 d-6}  \frac{\Phi_i}{T_i}=0\,, \qquad  \sum_{i=1}^{2 d-6}  \frac{1}{T_i}=0
\end{equation}

\subsubsection{Reissner-Nordstrom-AdS Black Hole in $d\ge 5$}
The asymptotically AdS static charged black hole in $d$-dimensions is described in detail in \cite{Zhang:2014jfa,Chamblin:1999tk}
with a radii function,
\begin{equation}
\Delta(r)=1-\frac{2 \mu}{r^{d-3}}+\frac{q^2}{r^{2 (d-3)}}+\frac{2 r^2}{(d-2)(d-1) L^2}
\end{equation}
where \(\mu\) and \(q\) have the same definitions of (\ref{eq:defsRNHighDim}). In this case, Hawking's temperature, entropy and electric potential defined at the respective BH horizons located at \(r_i\) can be generalized,
\begin{equation}
T_{i}=\frac{\frac{2 r_i^{2d+2} (d-1)}{L^2} + (d-2)(d-3) r_i^{2 d}-\frac{32 \pi^2 Q^2 r_i^6}{ \mathcal{A}^2_{d-2}}}{4 \pi (d-2) \, r_i^{2d+1} } 
\end{equation}
\begin{equation}
S_{i}=\frac{\mathcal{A}_{d-2}\, \, r_{i}^{d-2} }{4}\,,\qquad \Phi_{i}=\sqrt{\frac{d-2}{2 (d-3)}} \frac{ Q}{r_{i}^{d-3}}\,.
\end{equation}
Note that the geometrical relation of thermodynamics  (\ref{firstlaw}) for the above quantities is satisfied as well as the following relations for \(d\ge4\)
\begin{eqnarray}
\sum_{i=1}^{2(d-2)} S_i &=&
\begin{cases}
0 &\text{d is odd}\\
\frac{1}{2} \left(\frac{2-d}{2} \right)^{(d-2)/2} \,\mathcal{A}_{d-2} \, L^{d-2}&\text{d is even}
\end{cases}
%
\end{eqnarray}
and
\begin{equation}
\sum_{i=1}^{2(d-2)} \frac{\Phi_i}{T_i}=0\,, \qquad  \sum_{i=1}^{2(d-2)} \frac{1}{T_i}=0\,.
\end{equation}

\subsubsection{Kerr-Newman Black Hole}

The metric of a four dimensional (asymptotically flat) rotating, charged black hole was found in \cite{Caldarelli:1999xj}. The function, that defines the horizons is
\bea
\Delta(r)=(r^2+a^2 ) - 2 M r + Q^2
\eea
when $\Delta(r_{\pm})=0$ and $r_{\pm}= M\pm \sqrt{M^2-(a^2+Q^2)}$. 
Besides the black hole mass $M$ and angular momentum $J=a M$, one can easily compute the entropy, Hawking's temperature, angular velocity and electric potential that are respectively given by
\bea
S_{\pm}=4 \pi \left(r_{\pm}^{2}+a^{2}\right) \,, \quad
T_{\pm}=\frac{r^2_{\pm}-(a^{2}+Q^{2})}{4 \pi r_{\pm} \left(r_{\pm}^{2}+a^{2}\right)} \,,\quad
\Omega_{\pm}= \frac{a}{a^{2}+r_{\pm}^{2}} \,, \quad
\Phi_{\pm}=\frac{Q \, r_{\pm} }{a^{2}+r_{\pm}^{2}}   \,.
\eea
Employing these quantities the geometrical law of thermodynamics (\ref{firstlaw}) and the following relations can be verified
\bea
\sum^2_{i=1} S_i =2 \pi \left( 2 M^2-Q^2 \right)\,,
\eea
and
\bea
\sum_{i=1}^2\frac{\Phi_i}{T_i}\ne0\,, \qquad  \sum_{i=1}^2\frac{\Omega_i}{T_i}=0\,, \qquad  \sum_{i=1}^2\frac{1}{T_i}\ne0\,.
\eea

\subsubsection{Kerr-Newman-AdS Black Hole}
The metric of a four dimensional rotating, charged black hole in AdS is presented in \cite{Caldarelli:1999xj}. The function
\bea
\Delta(r)=(r^2+a^2 )\left(1+\frac{r^2}{L^2} \right) - 2 M r + Q^2
\eea
defines the horizon radii located at $\Delta(r_{i})=0$ with $i=1,...,4$.  The mass, angular momentum, entropy, Hawking's temperature, angular velocity and electric potential parameters are respectively given by

\bea
E=\frac{M}{ \left(1-a L^{-2}\right)^2}  \,, \qquad
J=\frac{a M}{ \left(1-a L^{-2}\right)^2}  \,, \qquad
S_{i}= 4 \pi \frac{\left(r_i^{2}+a^{2}\right)}{\left(1-\frac{a^{2}}{L^{2}}\right)} \,, 
\eea
\bea
T_{i}=\frac{r_{i}\left(1+\frac{a^{2}}{L^{2}}+3 \frac{r_{i}^{2}}{L^{2}}-\frac{a^{2}+Q^{2}}{r_{i}^{2}}\right)}{4 \pi\left(r_{i}^{2}+a^{2}\right)}  \,, \qquad
\Omega_{i}= \frac{a\left(1-\frac{a^{2}}{L^{2}}\right)}{a^{2}+r_i^{2}}  \,, \qquad
\Phi_{i}=\frac{Q \, r_i }{a^{2}+r_i^{2}}  \,.
\eea
The geometrical law of thermodynamics (\ref{firstlaw}) for the above quantities is satisfied as well as the following relations
\bea
\sum^4_{i=1} S_i =- 8 \pi L^2\,,
\eea
and 
\bea
\sum_{i=1}^4\frac{\Phi_i}{T_i}=0\,, \qquad  \sum_{i=1}^4\frac{\Omega_i}{T_i}=0\,, \qquad  \sum_{i=1}^4\frac{1}{T_i}=0\,,
\eea

\subsubsection{Non-Extremal Rotating Black Holes in Minimal $d=5$ Gauged Supergravity}
The solution of a non-extremal rotating black hole in minimal $d=5$ gauged Supergravity
 is presented in \cite{Chong:2005hr}. The coordinate choice is a Boyer-Lindquist type of \((t, r, \theta, \phi, \psi)\).  The function
\bea
\Delta(r)=\frac{\left(r^{2}+a_1^{2}\right)\left(r^{2}+a_2^{2}\right)\left(1+L^{-2} r^{2}\right)+Q^{2}+2 a_1 a_2 Q}{r^{2}}-2 M
\eea
defines the horizon radii located at $\Delta(r_{i})=0$ with $i=1,...,6$. The entropy, Hawking's temperature, angular velocities and electric potential parameters are respectively given by
\bea \label{eq:relSUGRA}
S_{i}&=&\frac{\pi^{2}\left[\left(r_{i}^{2}+a_1^{2}\right)\left(r_{i}^{2}+a_2^{2}\right)+a_1 a_2 Q\right]}{2 \left(1-a_1^2 L^{-2}\right) \left(1-a_2^2 L^{-2}\right) r_{i}} \,,\\
T_{i}&=&\frac{r_{i}^{4}\left[\left(1+L^{-2}\left(2 r_{i}^{2}+a_1^{2}+a_2^{2}\right)\right]-(a_1 a_2+Q)^{2}\right.}{2 \pi r_{i}\left[\left(r_{i}^{2}+a_1^{2}\right)\left(r_{i}^{2}+a_2^{2}\right)+a_1 a_2 Q\right]}\,, \nonumber \\
\Omega_i^{(1)}&=&\frac{a_1 \left(r_{i}^{2}+a_2^{2}\right)\left(1+L^{-2} r_{i}^{2}\right)+a_2 Q}{\left(r_{i}^{2}+a_1^{2}\right)\left(r_{i}^{2}+a_2^{2}\right)+a_1 a_2 Q}  \,, \nonumber \\
\Omega_i^{(2)}&=& \frac{a_2 \left(r_{i}^{2}+a_1^{2}\right)\left(1+L^{-2} r_{i}^{2}\right)+a_1 Q}{\left(r_{i}^{2}+a_1^{2}\right)\left(r_{i}^{2}+a_2^{2}\right)+a_1 a_2 Q}  \,, \nonumber \\
\Phi_{i}&=& \left( \ell^{\mu} A_{\mu} \right)_i  \nonumber
\eea
where
\bea
A_i&=&\frac{\sqrt{3} Q}{r_i^{2}+a_1^{2} \cos ^{2} \theta+a_2^{2} \sin ^{2} \theta}\left(\frac{1-a_1^{2} L^{-2} \cos ^{2} \theta-a_2^{2} L^{-2} \sin ^{2} \theta }{\left(1-a_1^2 L^{-2}\right) \left(1-a_2^2 L^{-2}\right)}d t \right. \\
  &-&\left.\frac{a_1 \sin ^{2} \theta \, }{1-a_1^2 L^{-2}}d \phi- \frac{a_2 \cos ^{2} \theta \, }{1-a_2^2 L^{-2}}d \psi \right) \nonumber \\
\ell&=&\frac{\partial}{\partial t}+\Omega_{i}^{(1)} \frac{\partial}{\partial \phi}+\Omega_{i}^{(2)}  \frac{\partial}{\partial \psi} \nonumber 
\eea
The generalized first law of thermodynamics (\ref{firstlaw}) takes the form 
\begin{equation} \label{eq:firstSUGRA}
dE=T_i dS+\Omega_i^{(1)} dJ_{(1)}+\Omega_i^{(2)} dJ_{(2)}+\Phi_{i} dQ
\end{equation}
where the mass and angular momenta are defined respectively as
\begin{equation} \label{eq:massSUGRA}
E=\frac{\pi  L^4 \left[M \left(3 L^4-\left(a_1^2+a_2^2\right) L^2-a_1^2 a_2^2\right)-2 a_1 a_2 Q \left(a_1^2+a_2^2-2 L^2\right)\right]}{4
   \left(a_1^2-L^2\right){}^2 \left(L^2-a_2^2\right){}^2}
\end{equation}
\bea \label{eq:angSUGRA}
J_{(1)}=\frac{\pi\left[2 a_1 M+Q a_2\left(1+a_1^{2} L^{-2}\right)\right]}{4\left( 1-a_1^2 L^{-2}\right)^{2} \left( 1-a_2^2 L^{-2}\right)} \,\qquad
J_{(2)}=\frac{\pi\left[2 a_2 M+Q a_1\left(1+b^{2} L^{-2}\right)\right]}{4 \left( 1-a_2^2 L^{-2}\right)^{2} \left( 1-a_1^2 L^{-2}\right)}\,.
\eea
Our results show that the relation of the entropy sum vanishes
\bea
\sum^6_{i=1} S_i =0
\eea
and the relations between the intensive quantities yield
\bea
\sum_{i=1}^6\frac{\Phi_i}{T_i}=0\,,  \qquad  \sum_{i=1}^6\frac{\Omega_i^{(1,2)}}{T_i}=0\,,\qquad  \sum_{i=1}^6\frac{1}{T_i}=0\,.
\eea

\subsubsection{Charged Rotating Black Holes in $d=6$ Gauged Supergravity}
The metric of a six dimensional rotating, charged black hole in $d=6$ gauged supergravity is presented in \cite{Chow:2008ip}. The function
\bea
\Delta(r)=\left(r^{2}+a_1^{2}\right)\left(r^{2}+a_2^{2}\right)+L^{-2}\left[r\left(r^{2}+a_1^{2}\right)+Q\right]\left[r\left(r^{2}+a_2^{2}\right)+Q\right]-2 M r
\eea
defines the horizon radii located at $\Delta(r_{i})=0$ with $i=1,...,6$. The entropy, angular velocities, electric potential and Hawking's temperature are respectively given by
\bea
S_{i}&=&\frac{2 \pi^{2}\left[\left(r_{i}^{2}+a_1^{2}\right)\left(r_{i}^{2}+a_2^{2}\right)+Q r_{i}\right]}{3  \left(1-a_1^2 L^{-2}\right) \left(1-a_2^2 L^{-2}\right)} \,,\\
\Omega_i^{(1)}&=&\frac{a_1\left[\left(1+L^{-2} r_{i}^{2}\right)\left(r_{i}^{2}+a_2^{2}\right)+Q  r_{i} L^{-2}\right]}{\left(r_{i}^{2}+a_1^{2}\right)\left(r_{i}^{2}+a_2^{2}\right)+Q r_{i}} \,, \nonumber \\
\Omega_i^{(2)}&=& \frac{a_2\left[\left(1+L^{-2} r_{i}^{2}\right)\left(r_{i}^{2}+a_1^{2}\right)+Q  r_{i} L^{-2}\right]}{\left(r_{i}^{2}+a_1^{2}\right)\left(r_{i}^{2}+a_2^{2}\right)+Q r_{i}}  \,, \nonumber \\
\Phi_{i}&=&\frac{Q r_i}{\pi} \frac{ \left( 1-a_1^2 L^{-2}\right) \left(1-a_2^2 L^{-2}\right) }{\left(r_{i}^{2}+a_1^{2}\right)\left(r_{i}^{2}+a_2^{2}\right)+Q r_{i}} \nonumber \,,
\eea
and
\bea
T_{i}&=&\frac{2 \left(1+L^{-2} r_{i}^{2}\right) r_{i}^{2}\left(2 r_{i}^{2}+a_1^{2}+a_2^{2}\right)}{4 \pi r_{i}\left[\left(r_{i}^{2}+a_1^{2}\right)\left(r_{i}^{2}+a_2^{2}\right)+Q r_{i}\right]}\nonumber \\
&+&\frac{4 Q L^{-2} r_{i}^{3}-\left(1-L^{-2} r_{i}^{2}\right)\left(r_{i}^{2}+a_1^{2}\right)\left(r_{i}^{2}+a_2^{2}\right)-Q^{2} L^{-2}}{4 \pi r_{i}\left[\left(r_{i}^{2}+a_1^{2}\right)\left(r_{i}^{2}+a_2^{2}\right)+Q r_{i}\right]}
\eea
\begin{equation}
E=\frac{\pi \left[2 M\left(\frac{1}{ (1-a_1^2 L^{-2})}+\frac{1}{ (1-a_2^2 L^{-2})}\right)+Q\left(1+\frac{ (1-a_1^2 L^{-2})}{ (1-a_2^2 L^{-2})}+\frac{ (1-a_2^2 L^{-2})}{ (1-a_1^2 L^{-2})}\right)\right]}{3 (1-a_1^2 L^{-2}) (1-a_2^2 L^{-2}) }
\end{equation}
\begin{equation}J_{(1)}=\frac{ \pi  a_1 \left(2 M+(1-a_2 L^{-2}) Q\right)}{3 (1-a_1 L^{-2})^{2}(1-a_2 L^{-2})} \nonumber \end{equation}
\begin{equation}J_{(2)}=\frac{ \pi  a_2\left(2 M+(1-a_1 L^{-2}) Q\right)}{3 (1-a_1 L^{-2})(1-a_2 L^{-2})^{2}}\end{equation}
For the above quantities, the geometrical thermodynamical law (\ref{firstlaw}) is satisfied as well as the following relations
\bea
\sum^6_{i=1} S_i =0
\eea
and
\bea
\sum_{i=1}^6\frac{\Phi_i}{T_i}=0\,,  \qquad  \sum_{i=1}^6\frac{\Omega_i^{(j)}}{T_i}=0\,,\qquad  \sum_{i=1}^6\frac{1}{T_i}=0\,.
\eea

\section{General Thermodynamic Relations for Black Holes}
\label{sec:proof}

In this section we will show the new thermodynamic relations follow from the mechanical law of black hole horizons and the properties of the sum of the entropies.
Our starting point is the first law for all black hole horizons (\ref{firstlaw}) for a a black hole solution with $r_i$ horizons (including the event and Cauchy horizons). This fundamental property for all black hole horizons can be written as
\begin{equation}
dS_i=\frac{1}{T_i }dE-\sum_k \frac{\Omega^{(k)}_i}{T_i } \, dJ_{(k)}-\sum_l  \frac{ \Phi^l_i}{T_i}\, dQ_l\,,
\end{equation}
Now, adding all these equations together for every black hole horizons leads to the expression
\begin{equation}
 \sum_{i} d S_i = \left(\sum_i \frac{1}{T_i }\right)dE-\sum_a \left(\sum_i \frac{\Omega^{a}_i}{T_i }\right) \, dJ_a-\sum_b  \left(\sum_i \frac{ \Phi^b_i}{T_i}\right)\, dQ_b\,.
 \end{equation}
The left hand side is
\begin{equation}
 \sum_{i} d S_i = d\left( \sum_{i} S_i \right)\,,
 \end{equation}
such, that when $ \sum_{i} S_i $ is independent of the extensive quantities
\begin{equation}
 \sum_{i} S_i \ne f(E, J_a, Q_b)  \rightarrow \sum_i \frac{1}{T_i }=0\,,\,\, \sum_i \frac{\Omega^{a}_i}{T_i } =0 \,, \,\, \sum_i \frac{ \Phi^b_i}{T_i}=0\,.
 \end{equation}
Therefore, together with the universal property of a first law for every black hole horizon, it is only necessary to identify the functional dependence of the sum of the entropies of every horizon in the solution $\sum S_i $ to single out the thermodynamic relations the solution will obey.\\
For black hole solutions, such for the Schwarzschild, Kerr, Reissner-Nordstrom and Kerr-Newman black holes solutions, where $ \sum_{i} S_i = f(E,Q) $ we can infer that
\begin{equation}
 \sum_{i} S_i = f(E,Q)  \rightarrow \sum_i \frac{\Omega^{a}_i}{T_i } =0 \,.
 \end{equation}
while 
\begin{equation}
\sum_i \frac{1}{T_i }\ne 0\,, \qquad \sum_i \frac{ \Phi^b_i}{T_i}\ne 0\,.
 \end{equation}
%

\section{Discussion}
\label{sec:discussion}

We have verified that the equations of motion of quantum field theories in curved backgrounds, more precisely the the KG-equation in black hole backgrounds, encode fundamental black hole thermodynamic relations. The universality of the entropy variation relations was established for a large class of black holes, both in asymptotically flat and asymptotically Anti-de Sitter spacetimes in four and higher dimensions.
The monodromies capture the infinitesimal changes in the black hole background produced by the field excitations. This emergent link between monodromies and entropies results in a thermodynamic identity for the sum of all horizon entropy variations. This raises the possibility of further thermodynamic relations defined as independent sums of temperatures and angular velocities defined at every black hole horizon. The origin of these relations can be put down to the fact that $\sum_{i} S_i \ne f(E, J_k, Q_l)$. Our explicit results indicate that black hole solutions with more than one Cauchy horizon obey all the thermodynamics relations (\ref{conditions1}) and (\ref{conditions2}). We emphasize that the thermodynamic summation relations apply in theories involving multi-charge black holes, including black hole solutions in gauged supergravities. For example, the charged rotating black holes in minimal $d = 5$ and $d = 6$ gauged Supergravity are consistent with
\bea
\delta\left(\sum^{\mathcal{K}}_{i=1} S_i \right) =0
\eea
From the more formal perspective, we have here worked out the details for the link between the monodromies and entropies for charged black holes. Nevertheless, our thermodynamic analysis contains robust evidence to argue that similar results will be found. A way to verify formally this proposal, is to consider a field interacting with electromagnetism through the equations of motion for a massless charged scalar. We leave this analysis for the future.

Note that the KG-equation for extremal asymptotically flat black hole solutions contain two irregular singular points (the event horizon and infinity). This is in contrast with non-extremal asymptotically flat black hole solutions containing only one irregular singular point at $r=\infty$. The emergence of thermodynamic relations is also expected and will be studied elsewhere.

Black hole solutions in GR in higher dimensions also contain more exotic solution such as black rings, bicycling black rings, black branes and black strings. It would be interesting to analyze the first law of all black hole horizon, and the emergent thermodynamic relations from monodromies for all these other cases. Further tests will include for black hole solutions in alternatives of GR e.g. Gauss-Bonnet or $f(R)$-theories.

\section*{Acknowledgements}

We are grateful to Malcolm Perry for helpful discussions. This work was supported by the NSF grant PHY-1707571 at Utah State University. The work of MJR is partially supported through the grants SEV-2016-0597 and PGC2018-095976-B-C21 from MCIU/AEI/FEDER, UE.

\appendix
\appendix
%

\section{Regular and irregular singular points for black holes}
\label{app:IrregularSingPoints}

Here we elaborate some subtleties that arise in the computations of the fake monodromy $\lambda_{\infty}$. Section \ref{subsec:fuchs} focused on regular singular points. In order to determine the fake monodromy around an irregular singular point, such as $r=\infty$ in this paper, we 
consider an ODE of the form:
\bea\label{eq:ODE}
\frac{d^2 {R}}{dr^2}+p (r) \,\frac{d {R}}{dr}+q (r) \, {R}=0\,. \label{eq:A1}
\eea
Then the following definition can be applied to determine the nature of any singularity.
{\it Definition}: 
Point $a$ is an ordinary point when functions  $p(r)$ and $q(r)$ are analytic at $r=a$. When the functions $p(r), q(r)$ each have poles on e.g. $r=a$ we call a singular point $a$ to be {\it regular} if either $p(r)$ or $q(r)$ diverges as $r\rightarrow a$ but 
\bea
\lim_{r\rightarrow a} (r-a) \,\,p(r) = \text{finite}\ =p_0 \, , \qquad \lim_{r\rightarrow a} (r-a)^2\, \, q(r) = \text{finite}=q_0\,.
\eea
Otherwise, we call it {\it irregular}.

By performing a suitable coordinate transformation to a new variable $u(r)=f(r) R(r)$, for some function $f(r)$ to be determined, we may write this differential equation (\ref{eq:ODE}) in the language of \cite{Erdelyi:1956} as 
\bea\label{eq:Erdelyi}
 u'' + \tilde{q}(r) \, u= 0, \qquad \text{with} \qquad \tilde{q}(r)={q}_0 +\frac{{q}_1}{r} +\frac{{q}_2}{r^2} +... \,.
\eea
around the irregular singular point $ r=\infty$.
Using equation (3) in \cite{Erdelyi:1956} Section 3.2:
\bea
u(r)= e^{\pm\sqrt{q_0}r}r^{i\,\lambda_{\infty}}\Big{(}1 + \sum\limits_{n\geq1}\frac{a_n}{r^n} \Big{)}
\eea
where the fake monodromy is determined by $\lambda_{\infty} = \frac{\pm  q_1}{2i\,\sqrt{q_0}}$. This prescription for computing $\lambda_{\infty}$ is employed throughout this paper.
We present the explicit computations for the fake $\lambda_{\infty}$ (at the irregular singular point $r=\infty$) for a d-dimensional Schwarzschild black hole back to back with the asymptotically AdS black hole cousin (with all regular singular point, including $r=\infty$).\\

{\it Example: $d\ge4$ Schwarzschild Black Hole (asympt. flat and $AdS$)} \\

Consider first an AdS$_{d}$ Schwarzschild Black Hole:
\bea
ds^2&=& -\Delta \, dt^2 + \Delta^{-1}dr^2 +r^2 d\Omega \\
\text{where}\qquad{}&\Delta&=1-\frac{2M}{r^{d-3}} +\frac{r^2}{L^2}\,.
\eea
Then, 
\bea
g_{\mu\nu}&=& \begin{pmatrix}
-\Delta& &  \\
&\Delta^{-1}& \\
& & r^2 [\Sigma]_{ij}
\end{pmatrix}
\eea
Here, $[\Sigma]_{ij}$ is the surface submetric. By dimensional analysis, it is apparent that, if  $\dim \Sigma = n$: $\det \Sigma = r^{2n}f(\phi_\alpha)$, where $f(\phi_\alpha)$ is some function of the generalized angles (in fact, using generalized polar coordinates: $f(\phi_\alpha)=\prod\limits_{i=1}^{n-2}\sin^{n-1-i}(\phi_i)$). Because it is a strictly projected submetric, $n=d-2$; then, det$g=-r^{2d-4}f(\phi_\alpha)$. \\ \\
We now turn to the Klein-Gordon equation
\bea
K_g[\Phi]\equiv \frac{1}{\sqrt{-g}}\partial_{\mu} (\sqrt{-g} g^{\mu\nu} \partial_{\nu})[\Phi]=0\,,\\
\frac{1}{r^{d-2}\sqrt{f(\phi_\alpha)}}\partial_{\mu} (r^{d-2}\sqrt{f(\phi_\alpha)} g^{\mu\nu} \partial_{\nu})[\Phi]=0\,,
\eea
\bea
\frac{1}{r^{d-2}\sqrt{f(\phi_\alpha)}}\Big{(} \partial_t [ r^{d-2}\sqrt{f(\phi_\alpha)} (-\Delta^{-1} \partial_t)] &+& \partial_r [ r^{d-2}\sqrt{f(\phi_\alpha)} (\Delta \partial_r)] \\
\nonumber& +&\partial_i [ r^{d-2}\sqrt{f(\phi_\alpha)} (\Sigma^{ij} \partial_j)]\Big{)}[\Phi] =0
\eea
Thus, invoking separation of variables, let $\Phi(t,r,\phi_\alpha)= W(t) R(r) \prod\limits_{i=1}^{n-2}\theta_i(\phi_i)$, so that
\bea
\frac{K_g[\Phi]}{\Phi}&=&0\,,\\
\frac{-\Delta^{-1}\partial^2_t[W(t)]}{W(t)} + \frac{\partial_r[r^{d-2} \Delta \partial_r R] }{r^{d-2}R(r)}+ \frac{\partial_i(\sqrt{f(\theta_\alpha)}\Sigma^{ij}\partial_j\prod\limits_{i=1}^{n-2}\theta_i(\phi_i))}{\sqrt{f(\theta_\alpha)}\prod\limits_{i=1}^{n-2}\theta_i(\phi_i)}&=&0\,.
\eea
Then, imposing eigenvalues:
\bea
\frac{-\Delta^{-1}\partial^2_t[W(t)]}{W(t)} = -\omega^2 \Leftrightarrow W(t)&=& W_0 e^{\pm i \omega t}\\
\frac{1}{\sqrt{f(\theta_\alpha)}\prod\limits_{i=1}^{n-2}\theta_i(\phi_i)}\partial_i(\sqrt{f(\theta_\alpha)}\Sigma^{ij}\partial_j\prod\limits_{i=1}^{n-2}\theta_i(\phi_i))&=& \sum K_{L_i}
\eea
So that we may rewrite the radial anstatz as
\bea
\frac{\omega^2}{\Delta} + \frac{1}{r^{d-2}R(r)}\partial_r[r^{d-2} \Delta \partial_r R] + \sum K_{L_i}&=&0\,,\\
\text{Or,}\qquad{}  \Big{[}\frac{1}{r^{d-2}}\partial_r[r^{d-2} \Delta \partial_r R] +\frac{\omega^2}{\Delta} + \sum K_{L_i}\Big{]}R(r)&=&0\,.
\eea
The transform 
\bea\label{eq:transform}
R(r)=(r^{d-2}-2Mr)^{-1/2}u(r)
\eea 
eliminates the $1^{\text{st}}$ order term, yielding
\bea
\partial_r^2u(r) + \Bigg{(}\frac{\partial_r(r^{d-2}\Delta \partial_r(r^{d-2}-2Mr)^{-1/2})}{r^{d-2}\Delta(r^{d-2}-2Mr)^{-1/2}} +\frac{\omega^2}{\Delta^2}\Bigg{)}u(r)=0\,.
\eea
In the above form, the asymptotic waveform is more apparent, albeit with a local, functional frequency $\partial_r^2u(r) + \bar{\omega}^2(r)u(r)=0$. Differentiating and expanding $\bar{\omega}^2$
\bea
 \bar{\omega}^2 = &-&\frac{\partial_r (r^{d-2}\Delta)\Big{(}(d-2)r^{d-3}-2M \Big{)}}{2r^{d-2} \Delta (r^{d-2}-2Mr)} \\
\nonumber&&-\frac{(d-2)(d-3)(r^{d-2}-2Mr)r^{d-4}-3\Big{(}(d-2)r^{d-3}-2M\Big{)}^2}{2(r^{d-2}-2Mr)^2}+ \frac{\omega^2}{\Delta^2}
\eea
Or
\bea
\nonumber\bar{\omega}^2 = &-&\frac{(\frac{d}{L^2}+\frac{d-2}{r^2}-\frac{2M}{r^{d-1}})\Big{(}(d-2)-\frac{2M}{r^{d-3}} \Big{)}}{2r^{2} (\frac{1}{L^2}+\frac{1}{r^2}-\frac{2M}{r^{d-1}}) (1-\frac{2M}{r^{d-3}})} \\
\nonumber&&-\frac{(d-2)(d-3)(1-\frac{2M}{r^{d-3}})-3\Big{(}(d-2)-\frac{2M}{r^{d-3}}\Big{)}^2}{2r^2(1-\frac{2M}{r^{d-3}})^2}\\
&&\qquad{}+ \frac{\omega^2}{r^{4}(\frac{1}{L^2}+\frac{1}{r^2}- \frac{2M}{r^{d-1}})^2}\label{eq:A2}
\eea
For AdS (or, finite $L$), asymptotically expanding at $r\rightarrow\infty$
\bea
\nonumber\bar{\omega}^2 \approx &-&\frac{L^2}{2r^2}\Big{(}\frac{d}{L^2}+\frac{d-2}{r^2}-\frac{2M}{r^{d-1}}\Big{)}\Big{(}(d-2)-\frac{2M}{r^{d-3}} \Big{)}\Big{[}\sum\limits_{n=0}\Big{(}\frac{r^{d-3}-2M}{r^{d-1}}\Big{)}^{n}l^{2n}\Big{]}\Big{[}\sum\limits_{n=0}\Big{(}\frac{2M}{r^{d-3}}\Big{)}^{n}\Big{]}\\
\nonumber&&-\frac{1}{2r^2}\Big{[}(d-2)(d-3)(1-\frac{2M}{r^{d-3}})-3\Big{(}(d-2)-\frac{2M}{r^{d-3}}\Big{)}^2\Big{]}\Big{[} \sum\limits_{n=1}n\Big{(}\frac{2M}{r^{d-3}}\Big{)}^{n-1}\Big{]}\\
&&\qquad{}+\frac{L^4\omega^2}{r^4}\Big{[} \sum\limits_{n=1}n\Big{(}\frac{r^{d-3}-2M}{r^{d-1}}\Big{)}^{n-1}L^{2n-2}\Big{]}\\
\approx &&\frac{L^4\omega^2}{r^4}\Big{(}1+O\big{(}\frac{1}{r^{2}}\big{)}\Big{ )}+\frac{(d-2)(d-3)}{2r^2}\Big{(} 1+ O\big{(}\frac{1}{r}\big{)}\Big{)}
\eea
Thus, using (\ref{eq:A1}), we see that $p_0 = 0$, and $q_0=\frac{(d-2)(d-3)}{2}$. Therefore, $r\to \infty$ is a regular singular point in AdS$_d$, $\forall d$. \\

Returning to (\ref{eq:A2}) and considering the flat case, $\lim\limits_{L\to\infty}\bar{\omega}^2=\bar{\omega}_{\text{flat}}^2$:

\bea
\bar{\omega}_{\text{flat}}^2&=&\frac{\omega^2}{(1- \frac{2M}{r^{d-3}})^2} -\frac{(d-2)(d-3)(1-\frac{2M}{r^{d-3}})-2\Big{(}(d-2)-\frac{2M}{r^{d-3}}\Big{)}^2}{2r^2(1-\frac{2M}{r^{d-3}})^2}\\
\nonumber&\approx& \Big{[}\omega^2-\frac{1}{2r^2}\Big{(}(d-2)(d-3)(1-\frac{2M}{r^{d-3}})-2\Big{(}(d-2)-\frac{2M}{r^{d-3}}\Big{)}^2 \Big{)}\Big{]}\sum\limits_{n=1}n\Big{(}\frac{2M}{r^{d-3}}\Big{)}^{n-1}
\eea
Clearly, due to the constant term, $\bar{\omega}^2$ fails the criteria in (\ref{eq:A1}): $\lim\limits_{r\to\infty}r^2\bar{\omega}^2=\infty$. Thus, we must analyze the point $r \to \infty$ as irregular.
As described in \cite{Erdelyi:1956}, the ODE is of the form (\ref{eq:Erdelyi}). 
%
Here $q_0=-\omega^2$ $\forall d$. For $d=4$, $q_1=4\omega^2 M^2$, while $\forall d\neq 4$, $q_1=0$. Using \cite{Erdelyi:1956} Section 3.2 equation (6), we find, 
\bea
\tilde{\lambda}_{\infty}^{(d=4)}=\pm 2M \omega\,,\qquad \tilde{\lambda}_{\infty}^{d>4}=0\,.
\eea
Then, in the $u(r)$ frame we find the asymptotic solutions to be
\bea
\nonumber u(r) &=& e^{\pm i \omega r }\cdot r^{\pm i \tilde{\lambda}_\infty}\Big{(} 1 + O \Big{(}\frac{1}{r} \Big{)}\Big{)}
\eea
Equivalently, we can determine the asymptotic expansion in the $R$ frame employing the transform (\ref{eq:transform}) between the different frames such that
\bea 
R(r)=e^{\pm i \omega r }\cdot r^{\pm i \tilde{\lambda}_\infty-(d-2)/2}\Big{(} 1 + O \Big{(}\frac{1}{r} \Big{)}\Big{)}
\eea 
In the case of $d=4$, we find
\bea
R(r)=e^{\pm i \omega r }\cdot r^{\pm 2 i M\omega-1}\Big{(} 1 + O \Big{(}\frac{1}{r} \Big{)}\Big{)}
\eea
Note that in the R-frame, one finds two (rather than one) values of the fake monodromy $\pm \tilde{\lambda}_\infty$. As we explain in Section \ref{sec:Overview} the frame which simplifies the identification between the monodromies and entropy variations is a frame in which one of the monodromies vanishes at each of the singularities. This complication is an artifact of the coordinate system, where having two monodromies at each point may result in ambiguities. We argue that this complication does not arise in the $\tilde{R}$-frame where the ODE is of the form (\ref{eq:FuchsGen}) with (\ref{eq:FuchsGenPQ}). Moreover, for consistency in implementing the Fuchs-type relation the monodormies have to be determined in the same frame. The $\tilde{R}$-frame was the preferred choice in the previous sections to find the monodromies $\alpha_i$ at the finite regular singular points. Inverting the relation (\ref{frame})
\bea
 \tilde{R}(r) =\prod_{i=1}^{\bar{\mathcal{K}}-1} (r-r_i)^{+i\tilde{\alpha_i}} {R}(r)\,,\qquad
%
\text{with} \qquad
%
\tilde{\alpha_i}
=\frac{\omega}{\bar{\Delta}'(r_i)}\,,
\eea
To be explicit, the associated asymptotic expansions  at $r=\infty$ in the $\tilde{R}$-frame 
\bea
\tilde{R}(r)&=&e^{- i \omega r }\cdot r^{-(d-2)/2}\Big{(} 1 + O \Big{(}\frac{1}{r} \Big{)}\Big{)}\,,\\
 \tilde{R}(r)&=&e^{+ i \omega r }\cdot r^{+i 2\tilde{\lambda}_\infty-(d-2)/2}\Big{(} 1 + O \Big{(}\frac{1}{r} \Big{)}\Big{)}
\eea
Within this frame, it is $\lambda_{\infty}=2\tilde{\lambda}_\infty$ the monodromy that will contribute to the Fuchs-type relation at $r=\infty$, and therefore the entropy variation sum. To summarize, the fake monodromies in the $\tilde{R}$ frame are
\bea
&&\text{For Asympt. Flat}_4: \lambda_\infty\rightarrow 4 M \omega\,, \\
&&\text{For Asympt. Flat}_{d>4}: \lambda_\infty\rightarrow 0\,.
\eea
%

%
\section{Properties of Kerr-AdS Horizons}
\label{app:PropertiesAdSKerr}

We briefly describe a few useful properties of the Kerr-AdS black hole solution:\\
1) The function  $\Delta(r)=0$ has four roots $r_1=r_+,r_2=r_-,r_3,r_4$ and can be written as 
\bea
\Delta(r)=(r-r_+)(r-r_-)(r-r_3)(r-r_4)/L^2\,.
\eea
2) The roots of $\Delta(r)=0$ are related in the following way
\bea
&&\sum_{i=1}^4 r_i=0\,,\qquad \prod_{i=1}^4 r_i =L^2 a^2\,,\\
&& \prod (r_i^2+a^2)= (2 M a L^2)^2 \,,\qquad \prod (1+r_i^2/L^2)= (2 M /L)^2\,.
\eea
3) This implies that we can re-write the parameters as
\bea
M=\frac{\left(L^2+r_-^2\right) \left(L^2+r_+^2\right) (r_-+r_+)}{2 L^2 \left(L^2-r_- r_+\right)}\,,\qquad a^2=\frac{r_- r_+ \left(L^2+r_-^2+r_- r_++r_+^2\right)}{L^2-r_- r_+}\,,\\
r_3=-\frac{1}{2} \left(r_- +r_++\sqrt{\frac{4 L^4+L^2 \left(3 r_-^2+2 r_- r_++3 r_+^2\right)+r_- r_+ (r_-+r_+)^2}{r_- r_+ - L^2}}\right)\,,\\
r_4=-\frac{1}{2} \left(r_- +r_+-\sqrt{\frac{4 L^4+L^2 \left(3 r_-^2+2 r_- r_++3 r_+^2\right)+r_- r_+ (r_-+r_+)^2}{r_- r_+ - L^2}}\right)\,,
\eea

\end{document}